\documentclass[twocolumn,superscriptaddress,amsmath,amssymb,aps,prl,floatfix]{revtex4-2}
\usepackage[usenames,dvipsnames]{color}

\usepackage{comment}
\usepackage{cancel}
\usepackage{ulem}
\usepackage{physics}
\usepackage{graphicx}
\usepackage{dcolumn}
\usepackage{bm}
\usepackage{gensymb}
\usepackage{tensor}
\usepackage{rotating}
\usepackage[percent]{overpic}
\usepackage{multibib}
\usepackage[breaklinks=true,colorlinks,citecolor=blue,linkcolor=blue,urlcolor=blue]{hyperref}
\footnotetext{These authors contributed equally to this work.}
\footnotetext{francesco.cioni@sns.it}

\begin{document}
\title{Simulating the Haldane model in ultra-clean GaAs heterostructures}
\date{\today}
\author{Francesco Cioni$^\star$${}^\ddag$}
\affiliation{NEST, Scuola Normale Superiore, I-56126 Pisa,~Italy}
\author{Lorenzo Cavicchi$^\star$}
\affiliation{Scuola Normale Superiore, Piazza dei Cavalieri 7, I-56126 Pisa,~Italy}
\author{Fabio Taddei}
\affiliation{NEST, Istituto Nanoscienze-CNR, Piazza S. Silvestro 12, I-56126 Pisa,~Italy}
\author{Marco Polini}
\affiliation{Dipartimento di Fisica dell'Universit\`a di Pisa, Largo Bruno Pontecorvo 3, I-56127 Pisa,~Italy}

\begin{abstract}
    The Haldane model represents the minimal lattice-based realization of a Chern insulator, exhibiting a quantized Hall conductance in the absence of Landau levels. Despite its conceptual elegance, the implementation in crystalline solids of the requisite pattern of Peierls phases breaking time-reversal symmetry remains experimentally demanding. In this work, we theoretically investigate the possibility to simulate the Haldane model in ultra-clean GaAs/AlGaAs heterostructures. Our proposal relies on recent experiments in which a high-mobility two-dimensional electron gas is subject to a gate-defined honeycomb electrostatic potential and a laterally periodic magnetic field generated by patterned ferromagnetic structures. The combined electrostatic and magnetic superlattices furnish a viable route to emulate the topological properties of the Haldane model.
\end{abstract}
\maketitle
{\color{blue} \it Introduction.---}More than two decades ago, Haldane proposed a minimal lattice model to prove that a quantized Hall response can exist {\it without} Landau levels~\cite{Haldane1988}. In its canonical formulation on the honeycomb lattice, the model realizes a Chern insulator by combining real nearest-neighbor hopping with complex next-nearest-neighbor tunneling phases that break time-reversal symmetry while maintaining {\it zero net magnetic flux} per unit cell (UC)~\cite{Haldane1988}.

While the Haldane model has been realized experimentally by using cold fermionic gases loaded in honeycomb optical lattices~\cite{Jotzu_Nature_2014}, its implementation in a crystalline solid is notoriously challenging~\cite{Qi2011,Wright_2013,Kim2017,Ho2017}. The essential requirement is a pattern of effective Peierls phases that breaks time-reversal symmetry on the scale of a UC while keeping the average flux per UC at or close to zero. This condition is generally extremely hard to realize experimentally in a solid-state platform. Consequently, many solid-state realizations of quantized Hall responses at zero applied fields have utilized different microscopic routes, most notably the quantum anomalous Hall (QAH) effect~\cite{nagaosa_2010} in magnetically-doped topological-insulator thin films~\cite{Yu2010Science,Chang2013,Chang2015NatMater,Deng2020Science} and multilayer structures consisting of alternating magnetic and undoped topological insulator layers~\cite{zhao_2020,zhao_2023}. More recently, moir\'e materials have demonstrated that Chern insulating states can emerge in electrically tunable minibands~\cite{Sharpe2019Science,Serlin2020,Chen2020Nature,Zhao2024}, but this typically relies on an emergent band reconstruction driven by strong electron-electron interactions.

Ultra-clean GaAs/AlGaAs heterostructures~\cite{Manfra2014,Chung2021} offer a complementary route in which the kinetic term of a target lattice Hamiltonian can be imposed externally on an exceptionally low-disorder two-dimensional electron gas (2DEG). A key milestone was the development of so-called``artificial graphene''~\cite{Park2009,Gibertini_PRB_2009,AG_review} in GaAs quantum wells, where a nanofabricated honeycomb modulation~\cite{Soibel1996,DeSimoni2010,Singha2011,Scarabelli2015,Wang2016,Wang2018,Du2021} produces minibands with graphene-like massless-Dirac-fermion features  and enables the exploration of interaction physics in a controllable lattice environment. The central advantage of the GaAs/AlGaAs heterostructure platform is that both geometry and energy scales are engineered at the device level while retaining direct access to standard transport probes.

More recently~\cite{Wang2024AEC}, gate-defined artificial electrostatic crystals in ultra-shallow GaAs quantum wells have pushed this approach toward Hamiltonian engineering. By continuously tuning the modulation strength and carrier density and using DC transport as a probing tool, Wang et al.~\cite{Wang2024AEC} demonstrated the formation of artificial band structures that can be reshaped in situ. Different regimes consistent with graphene-like linear (Dirac) dispersion and Kagome-like flat-band physics were explored within a {\it single} device. This level of tunability is difficult to achieve in etched lattices and is particularly relevant for solid-state quantum simulation, where resolving small gaps requires ultra-low disorder.

\begin{figure*}
    \centering
    \begin{overpic}[width=0.333\linewidth]{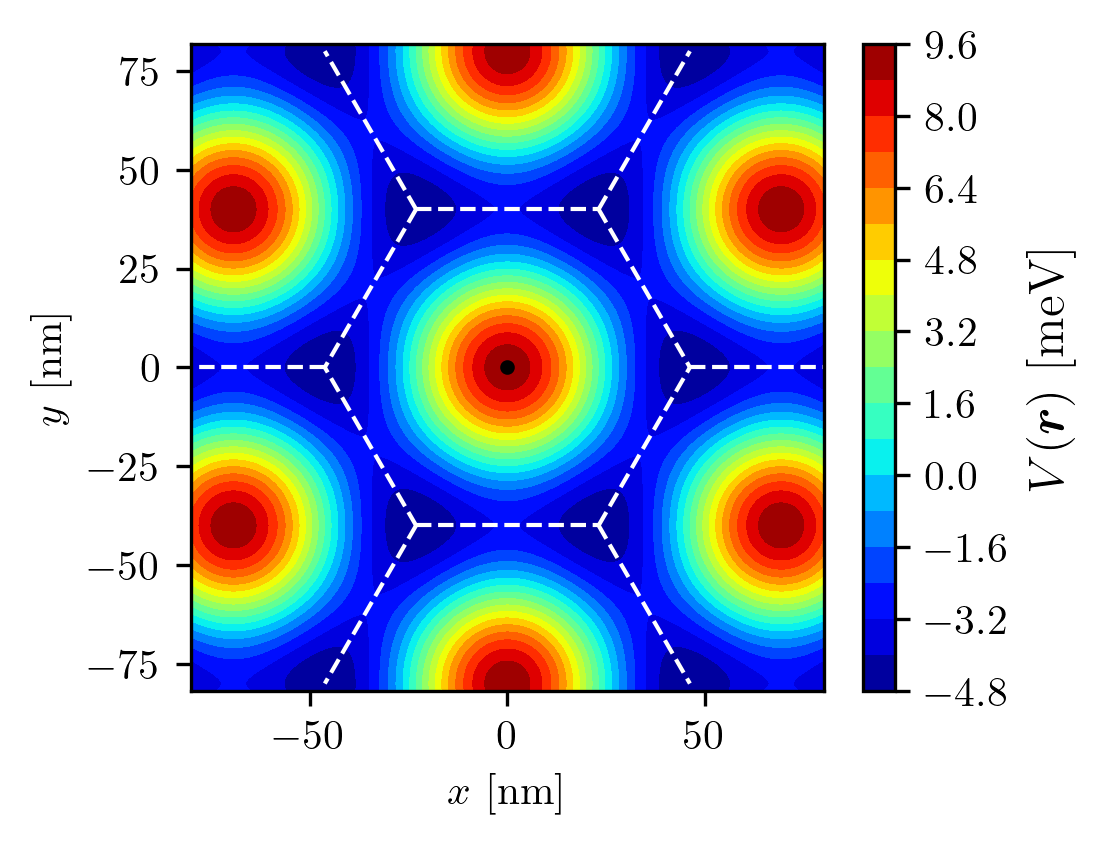}
        \put(0,72){(a)}
    \end{overpic}%
    \begin{overpic}[width=0.333\linewidth]{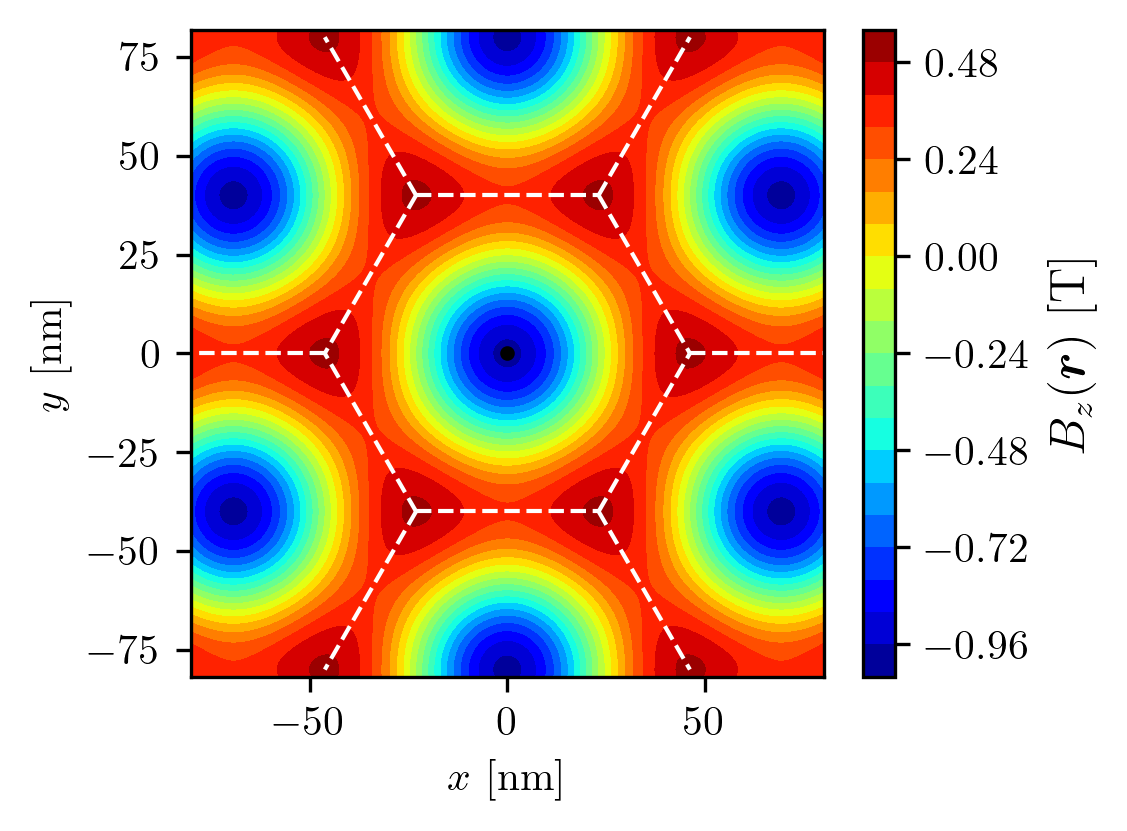}
        \put(0,72){(b)}
    \end{overpic}%
    \begin{overpic}[width=0.333\linewidth]{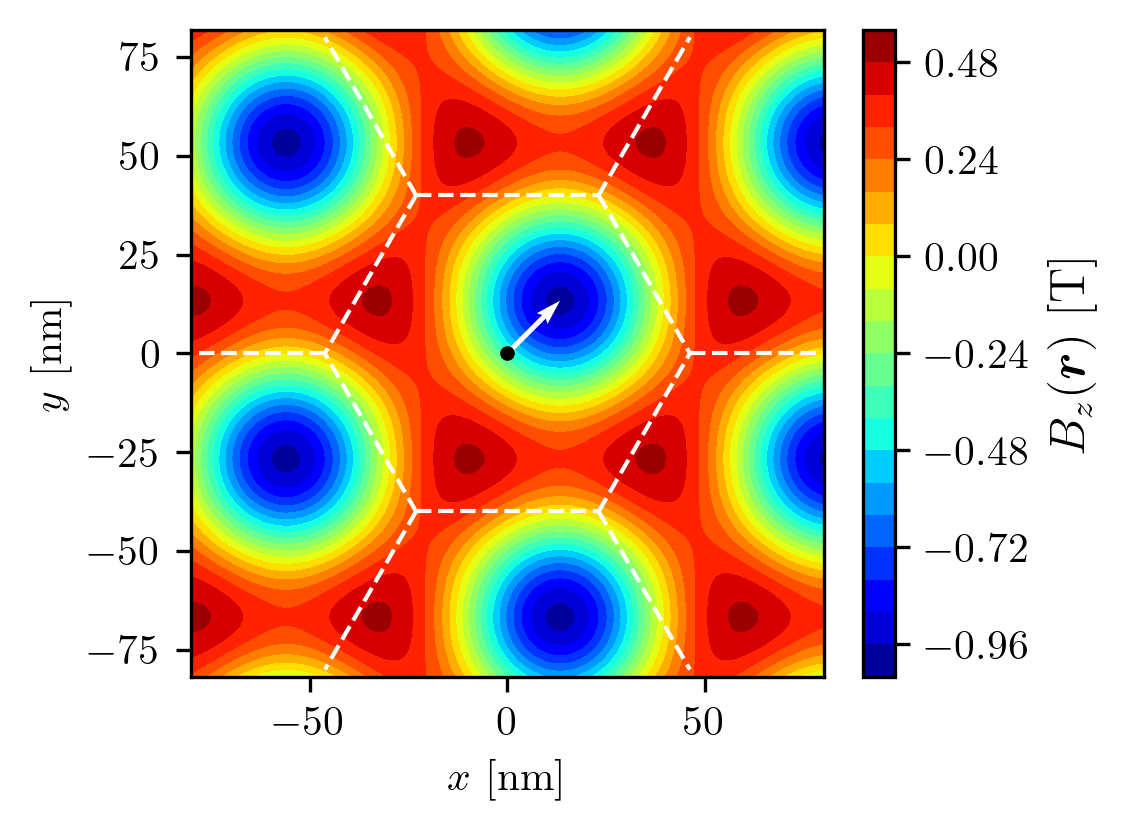}
        \put(0,72){(c)}
        \put(50,45){\color{white} $\bm r_0$}
    \end{overpic}
    \caption{Panel (a) Crystal field $V(\bm r)$ as obtained from Eq.~\eqref{V_crystal}. Panel (b) PMF $B_{z}(\bm r)$ as obtained from Eq.~\eqref{B_crystal}. Panel (c) PMF shifted by a vector $\bm r_0$ with respect to the crystal field, i.e.~$B_{z}(\bm r - \bm r_0)$. Parameters are fixed to be $a_0 = 80~{\rm nm}$, $W = 1.57~{\rm meV}$, and $\beta\Phi_0/2\pi = 330~{\rm mT}$. In all panels, the black dot in the center of the figure refers to the origin of the axis and is added as a guide for the eye.}
    \label{fig0}
\end{figure*}

Not only can the honeycomb lattice be engineered lithographically~\cite{Singha2011,Scarabelli2015,Wang2016,Wang2018} or electrostatically~\cite{Soibel1996,Park2009,DeSimoni2010,Du2021,Wang2024AEC}, but a controlled breaking of time-reversal symmetry in the orbital sector can be achieved through periodic magnetic-field engineering. In lateral magnetic superlattices, patterned superconducting or ferromagnetic structures placed above the heterostructure generate a spatially modulated perpendicular field $B_z(\bm r)$ and, hence, a designed vector potential $\bm A(\bm r)$. A seminal demonstration~\cite{Carmona1995} used superconducting stripe arrays to form a lateral magnetic superlattice directly coupled to a high-mobility 2DEG. Complementarily, regular arrays of micromagnets~\cite{Ye1995PRL,Nogaret2010,engdahl2024} have been shown to induce periodic magnetic-field modulations at submicron scales in GaAs/AlGaAs heterostructures. When combined with an engineered honeycomb electrostatic potential, such periodic magnetic architectures could provide a physically direct route to break time-reversal symmetry and target flux textures with vanishing flux over a UC.
We stress that any unintended uniform magnetic‑field component generated by a periodic array of micromagnets can be fully compensated by an external magnetic field of matching strength applied in the opposite direction.

These capabilities make simulating the Haldane model in GaAs/AlGaAs heterostructures a concrete and experimentally relevant possibility.
In this work, we address the feasibility of realizing such a quantum simulation platform by studying an artificial honeycomb electrostatic lattice combined with a spatially periodic magnetic field (PMF). We show that the superimposed magnetic pattern can induce a non-trivial topology of the bands, and we map the emerging low-energy physics directly onto the Haldane model. The broader objective is to realize a solid-state quantum simulator for a paradigmatic Chern-insulating Hamiltonian---conceptually simple, yet microscopically elusive in natural crystals---in an experimentally-relevant platform that is tunable and electrically reconfigurable.

\begin{figure*}
    \centering
    \begin{overpic}
        [width=0.325\linewidth]{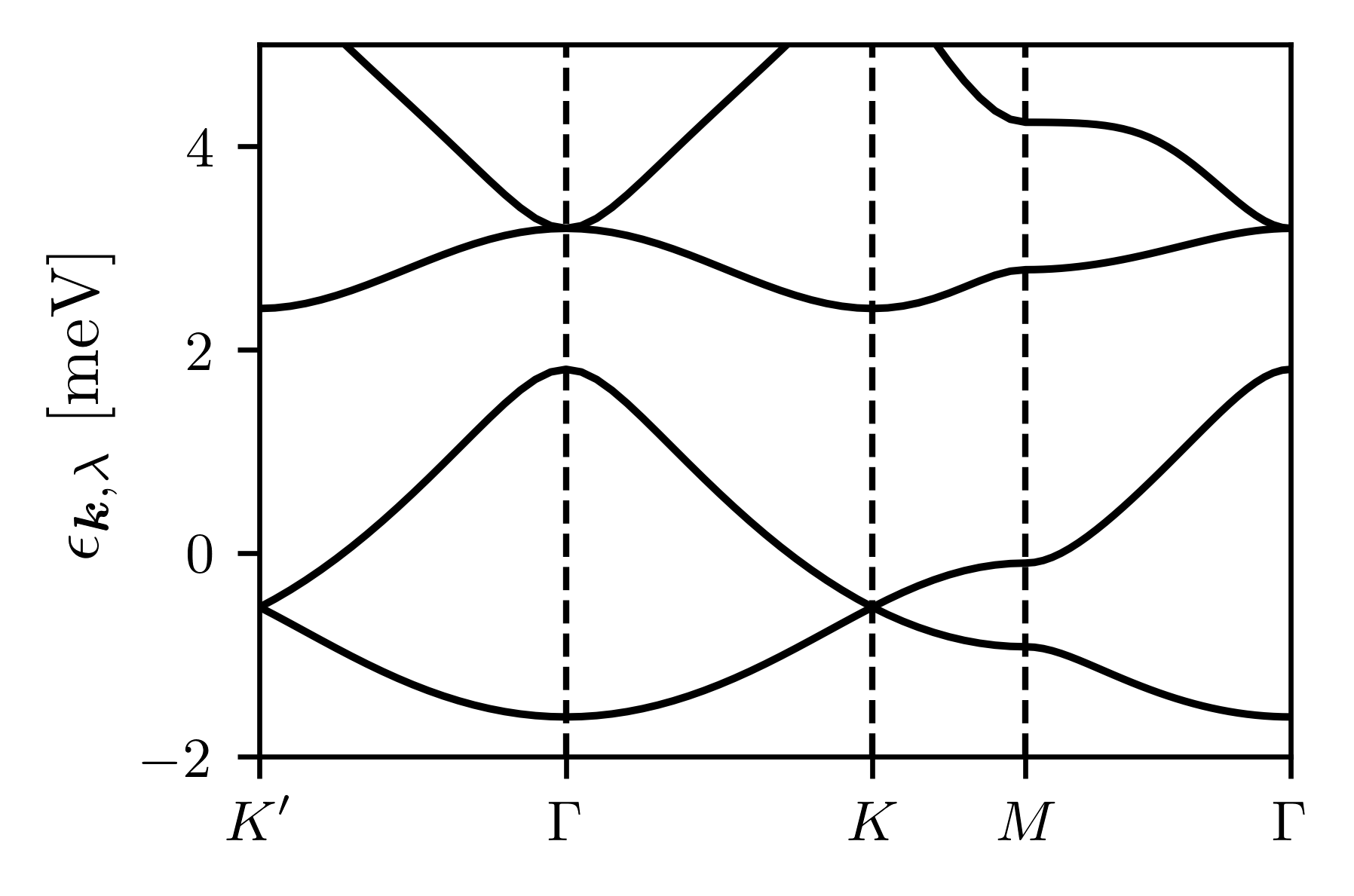}\put(0, 65){(a)}
    \end{overpic}
    \hfill
    \begin{overpic}
        [width=0.325\linewidth]{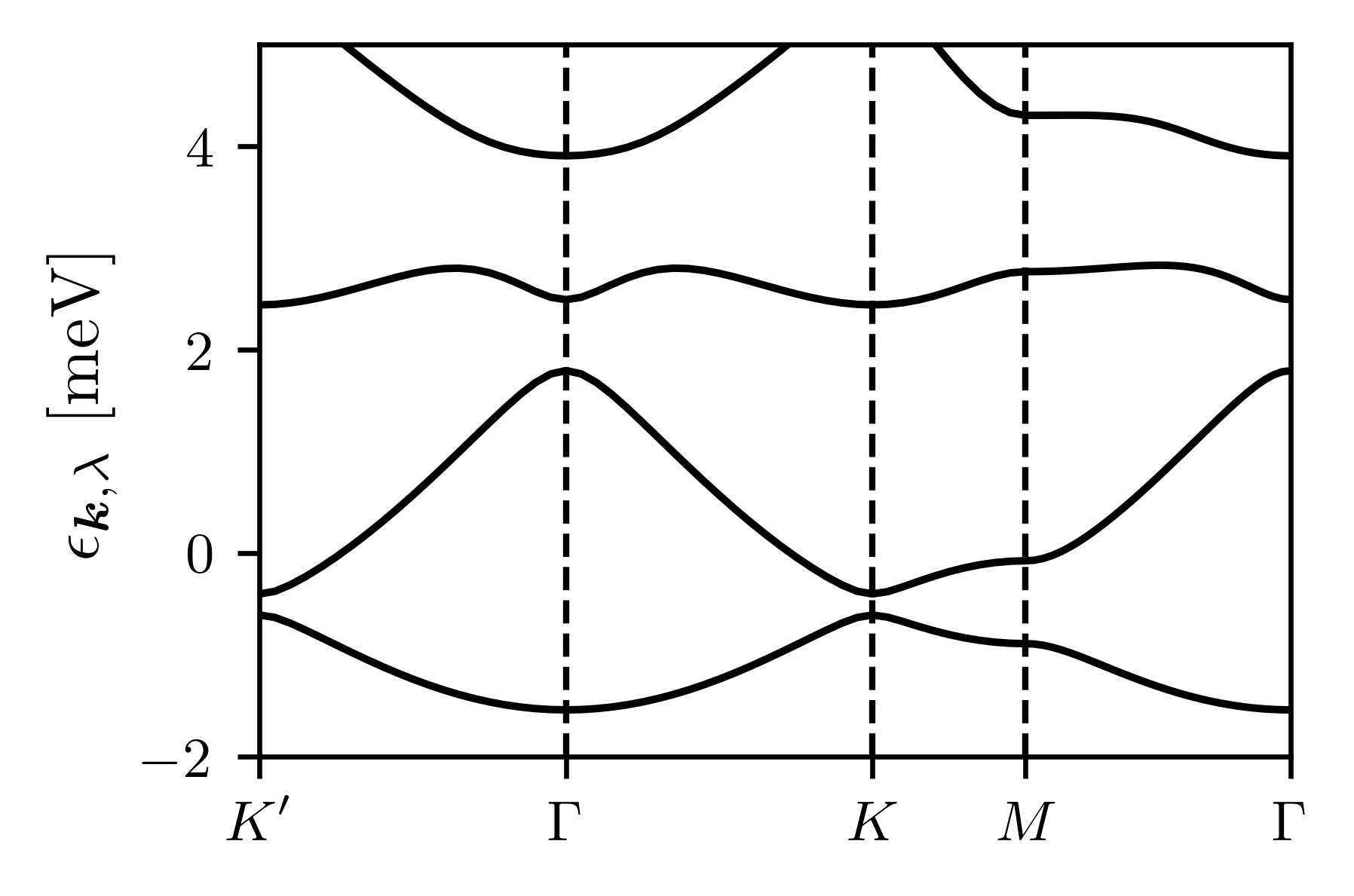}\put(0, 65){(b)}
    \end{overpic}
    \hfill
    \begin{overpic}
        [width=0.325\linewidth]{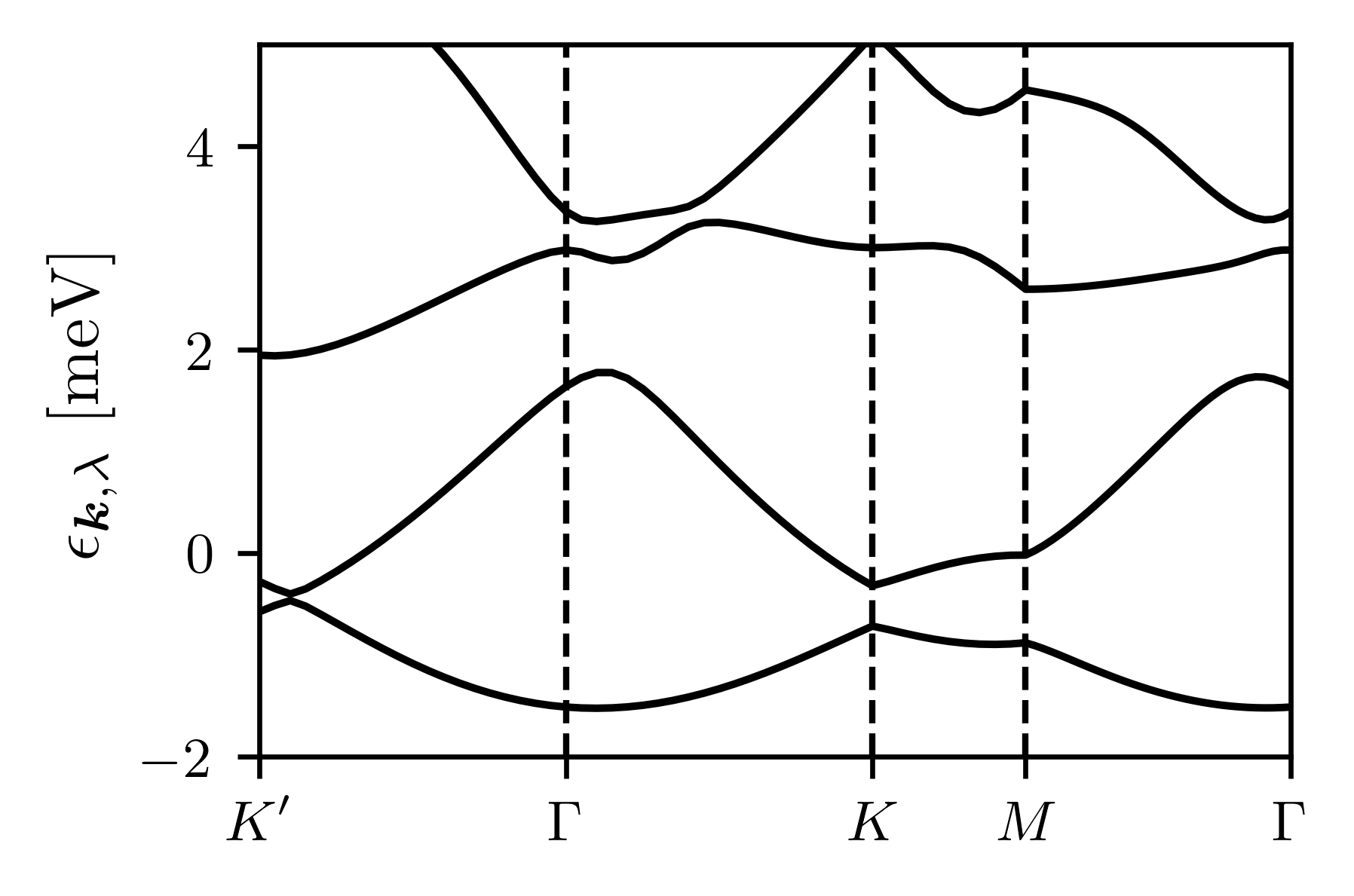}\put(0, 65){(c)}
    \end{overpic}
    \caption{Energy bands, along the high-symmetry path $K^\prime-\Gamma-K-M-\Gamma$ in the BZ, for different magnetic-field patterns. (a) $\beta \Phi_0 / 2\pi = 0$ and $\bm r_0=\bm 0$. (b) $\beta \Phi_0 / 2\pi = 330\ {\rm mT}$ and $\bm r_0=\bm 0$. (c) $\beta \Phi_0 / 2\pi = 330\ {\rm mT}$ and $\bm r_0=a_0(-0.45,-0.43)$. Other parameters are chosen as $a_0=80\ {\rm nm}$, $m^*=0.062\,m_e$, and $W=1.57\ {\rm meV}$.}
    \label{fig:Bands}
\end{figure*}

{\color{blue}\it Model Hamiltonian.---}
We consider a noninteracting two-dimensional system of spinless electrons subject to a crystalline potential $V(\hat{\bm r})$ and a PMF $\bm B(\hat{\bm r})$. Both fields are periodic with the same lattice periodicity. The magnetic field is chosen such that the net flux through each UC vanishes, which guaranties that the corresponding vector potential $\bm A(\hat{\bm r})$ can be taken to be periodic with the same period as the crystalline potential $V(\hat{\bm r})$. The single-particle Hamiltonian reads
\begin{equation}\label{eq:ham}
    \hat{\mathcal H}
    = \frac{1}{2m^*}\left[\hat{\bm p} + \frac{e}{c}\bm A(\hat{\bm r})\right]^2 + V(\hat{\bm r})~,
\end{equation}
where $m^*$ is the effective electron mass ($m^*\simeq 0.062~m_{\rm e}$ in GaAs~\cite{vianez_2023}, where $m_{\rm e}$ is the bare electron mass in vacuum), $-e<0$ is the elementary charge, and $c$ is the speed of light.

For concreteness, we specialize to a triangular lattice. The periodicity of $V(\bm r)$ and $\bm B(\bm r)$ is set by the primitive vectors
\begin{equation}\label{eqn:lattice_vectors}
    \bm a_{1/2} = a_0\left( \pm \frac{\sqrt{3}}{2}, \frac{1}{2}\right)~,
\end{equation}
with $a_0$ the lattice constant. Specifically, we take the crystal potential to be~\cite{Krix2022}
\begin{equation}\label{V_crystal}
    V(\bm r) = 2W \sum_{\ell=1,3,5} \cos\!\big(\bm G_\ell \cdot \bm r\big)~,
\end{equation}
and align the PMF along the
$\hat{\bm{z}}$ axis [$\bm{B}(\bm r)\equiv B_z(\bm r)\,\hat{\bm{z}}$], setting~\cite{Taillefumier2008}
\begin{equation}\label{B_crystal}
    B_z(\bm r)= \frac{\beta}{2\pi}\Phi_0 \sum_{\ell=1,3,5}
    \cos (\bm G_\ell \cdot \bm r )~.
\end{equation}
Here $W$ sets the strength of the crystalline potential, $\beta$ is a parameter--with the dimensions of an inverse area---controlling the PMF amplitude in units of $\Phi_0/(2\pi)$, $\Phi_0 = hc/e$ being the flux quantum, and
\begin{equation}
\bm G_\ell = \frac{4\pi}{\sqrt{3}a_0}\Big(\cos(\ell\pi/3),\,\sin(\ell\pi/3)\Big)
\end{equation}
are reciprocal lattice vectors. Since the magnetic field is periodic and carries zero flux per UC, we adopt a periodic (Coulomb) gauge in which $\bm\nabla\!\cdot\!\bm A(\bm r)=0$ and
\begin{equation}\label{eq: A_crystal}
    \bm A(\bm r) =
    \frac{\beta\Phi_0}{2\pi}\sum_{\ell=1,3,5}
    \frac{\bm G_\ell \wedge \hat{\bm z}}{\vert \bm G_\ell\vert ^2}
    \sin (\bm G_\ell\cdot\bm r)~.
\end{equation}
Fig.~\ref{fig0} displays the spatial profiles of $V(\bm r)$ and $B_z(\bm r)$ introduced in Eqs.~\eqref{V_crystal} and~\eqref{B_crystal}. Panel (a) of Fig.~\ref{fig0} shows that the crystal field generates an array of attractive wells arranged on a $D_{6h}$-symmetric honeycomb lattice. The PMF configuration depicted in panel (b) yields a similar spatial structure. In panel (c), the PMF is displaced by a vector $\bm r_0$ relative to the crystal field, i.e.~$B_{z}(\bm r - \bm r_0)$, which in general explicitly breaks both rotational and inversion symmetries of the $D_{6h}$ point group. The consequences of the rigid relative shift $\bm r_0=(x_0,y_0)$ between the crystal potential and the magnetic-field pattern are studied in depth in the next section. To produce the data in Fig.~\ref{fig0}, we choose $a_0=80\ {\rm nm}$, $m^*=0.062\,m_{\rm e}$, and $W=1.57\ {\rm meV}$, values compatible with experimentally accessible parameters in GaAs heterostructures~\cite{Wang2024AEC,Krix2022}. Unless indicated otherwise, we will employ this parameter set throughout the remainder of this work.

{\color{blue}\it Symmetries.---}
For any shift vector $\bm r_0$, the Hamiltonian is invariant under lattice translations. Indeed, the equalities $V(\bm r+\bm a)=V(\bm r)$ and $\bm A(\bm r+\bm a)=\bm A(\bm r)$, where $\bm a=i\bm a_1+j\bm a_2$ is a Bravais-lattice translation vector, imply that $[\hat{\cal H},\hat T_{\bm a}]=0$, $\hat T_{\bm a}\equiv e^{-(i/\hbar)\hat{\bm p}\cdot\bm a}$ being the translation operator. This symmetry allows one to diagonalize $\hat{\cal H}$ on a hexagonal Wigner-Seitz Brillouin zone (BZ), even though a generic non-zero shift $\bm r_0\neq\bm 0$ of the PMF relative to the crystalline potential  $V(\bm r)$ breaks the $C_{6z}$ symmetry of the latter.
On the other hand, time-reversal symmetry, ${\cal T}$, is broken for any $\bm r_0$ when $\beta\neq 0$---i.e.~when a finite magnetic field is present.

Let us now  consider the parity symmetry operator ${\cal P}$, which acts on the PMF, shifted by a generic vector $\bm r_0$ with respect to $V(\bm r)$, as follows

\begin{equation}\label{eqn:B_parity}
    {\cal P}\,B_z(\hat{\bm r}-\bm r_0)\,{\cal P}^\dagger
    = B_z(-\hat{\bm r}-\bm r_0) = B_z(\hat{\bm r}+\bm r_0)~.
\end{equation}
The second equality comes from the explicit form of $B_z(\bm r)$ in Eq.~\eqref{B_crystal}.
Eq.~\eqref{eqn:B_parity} shows that parity symmetry trivially holds when $\bm r_0=\bm 0$. Nonetheless, ${\cal P}$ is not always broken when $\bm r_0\ne \bm 0$, as can be shown by considering the parity-preserving shift vector $\bm r_0^{\rm p}$ that satisfies
\begin{equation}\label{eqn:cond1}
    \bm G_1\cdot\bm r_0^{\rm p} = n\pi~,\qquad
    \bm G_3\cdot\bm r_0^{\rm p} = m\pi~,
\end{equation}
with $n,m\in\mathbb Z$.
This implies that
\begin{equation}
    \bm G_5\cdot\bm r_0^{\rm p}
    = -(\bm G_1+\bm G_3)\cdot\bm r_0^{\rm p}
    = -(n+m)\pi~.
\end{equation}
Using $\bm r_0 = \bm r_0^{\rm p}$ in the expression of the PMF of Eq.~(\ref{B_crystal}) gives
\begin{widetext}
\begin{equation}\label{eqn:B_r0_parity}
    B_z(\bm r-\bm r_0^{\rm p}) = \frac{\beta}{2\pi}\Phi_0
    \Big[(-1)^{n}\cos(\bm G_1\cdot\bm r)
    +(-1)^{m}\cos(\bm G_3\cdot\bm r)
    +(-1)^{n+m}\cos(\bm G_5\cdot\bm r)\Big]~,
\end{equation}
\end{widetext}
which is even under parity since it is a superposition of even functions of $\bm r$. In particular, if both $n$ and $m$ are even, the field is explicitly equal to the case $\bm r_0=\bm 0$, and parity holds trivially.
The conditions in Eq.~\eqref{eqn:cond1} are explicitly satisfied by
\begin{equation}\label{eqn:r0_parity1}
    \bm r_0^{\rm p} = \frac{a_0}{2}\left(-m\frac{\sqrt{3}}{2},\, n+\frac{m}{2}\right)~.
\end{equation}

By analogous arguments, the combined parity-time-reversal symmetry (${\cal P}{\cal T}$), which acts on the Hamiltonian~\eqref{eq:ham} as
\begin{equation}
\begin{split}
    {\cal T}{\cal P} \hat{\mathcal{H}}(\bm A,\bm r_0){\cal P}^\dagger{\cal T}^\dagger &= \frac{1}{2m^*}\left[\hat{\bm{p}} - \frac{e}{c}\bm{A}(\hat{\bm{r}}+\bm r_0)\right]^2 + V(\hat{\bm{r}}) \\
    &= \hat{\mathcal{H}}(-\bm A,-\bm r_0)~,
\end{split}
\end{equation}
requires
\begin{equation}
    B_z(\bm r-\bm r_0) = -B_z(\bm r+\bm r_0),~\forall~{\bm r}~.
\end{equation}
It is easy to check that the above equality is never satisfied. This means that there is no vector $\bm r_0$ for which the system is ${\cal P}{\cal T}$-symmetric.

{\color{blue}\it Band structure and topological phase diagram.---}
Using a plane-wave expansion (see Sec.~\ref{appA} of the Supplemental Material (SM)~\cite{SM}), one can diagonalize $\hat{\cal H}$ and obtain the band dispersion $\epsilon_{\bm k,\lambda}$, where $\lambda$ represents the band index ($\lambda=1$ denoting the lowest band). Fig.~\ref{fig:Bands} shows three representative band structures for different magnetic-field patterns. Fig.~\ref{fig:Bands}(a) corresponds to $\beta\Phi_0/(2\pi)=0$, where Dirac cones appear at $K$ and $K'$. This is expected because, for the choice of the crystal field in Eq.~\eqref{V_crystal}, the electrons experience attractive potential minima arranged on a honeycomb lattice [see Fig.~\ref{fig0}(a)]. As shown in Fig.~\ref{fig:Bands}(b), a finite magnetic field, $\beta\Phi_0/(2\pi)\simeq 330\ {\rm mT}$ (corresponding to $\beta=1\times 10^{-3}\ {\rm nm^{-2}}$), opens gaps at both $K$ and $K'$. Parity remains preserved in this case since $B_z(\bm r)$ is aligned with $V(\bm r)$,~i.e. $\bm r_0=\bm 0$. To break both time-reversal and parity symmetry, one must choose a shift vector $\bm r_0\notin \{\bm 0,\bm r_0^{\rm p}\}$, $\bm r_0^{\rm p}$ being given by Eq.~\eqref{eqn:r0_parity1}. Fig.~\ref{fig:Bands}(c) shows the bands for a representative shift $\bm r_0$ chosen within the UC. Asymmetric gaps open at the Dirac points, which are also displaced away from $K$ and $K'$ due to the broken $C_{3z}$ symmetry. The magnetic field is fixed as in Fig.~\ref{fig:Bands}(b).

\begin{figure}
    \centering
    \begin{overpic}[width=0.8\columnwidth]{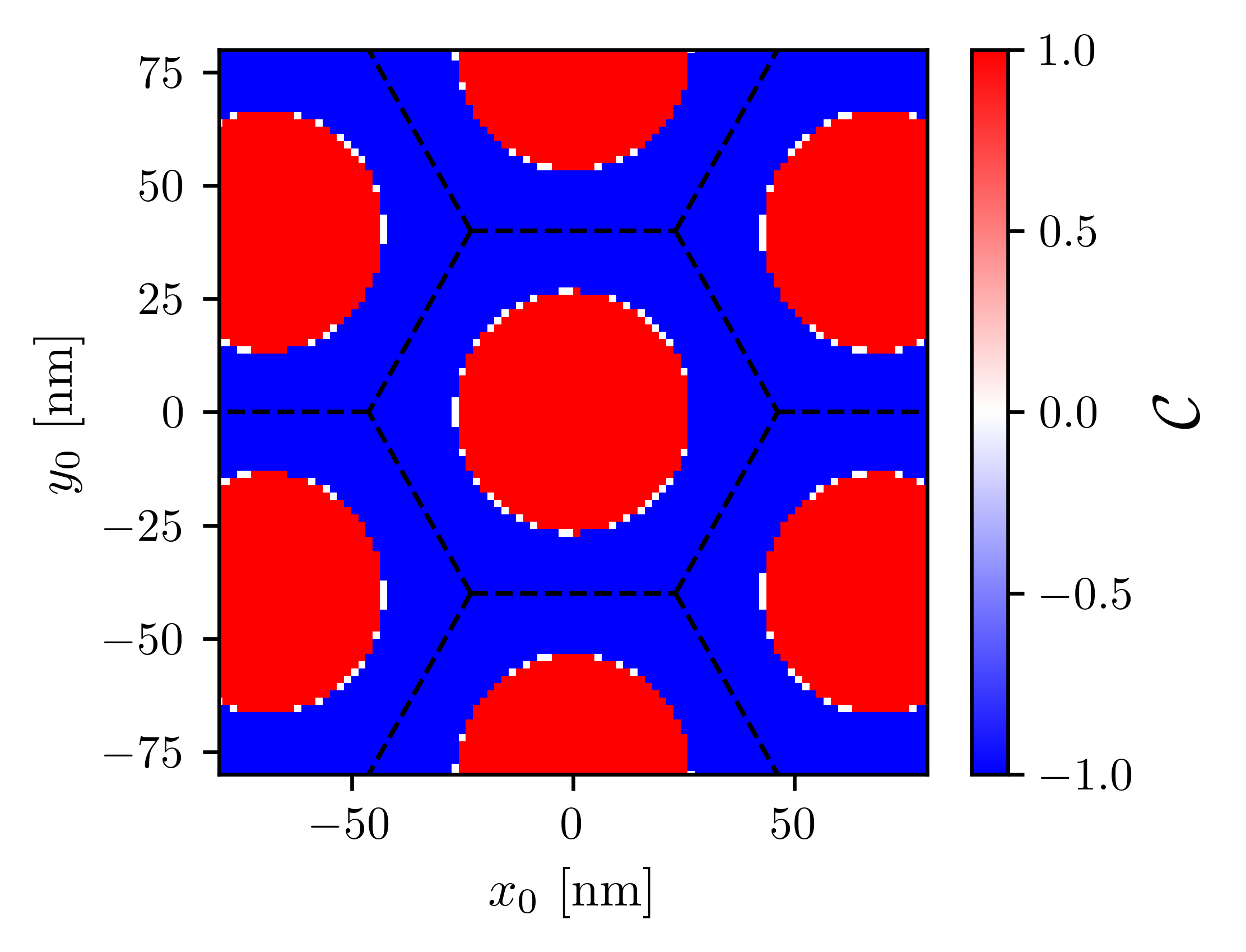}
        \put(0,71){(a)}
    \end{overpic}
    \begin{overpic}[width=0.8\columnwidth]{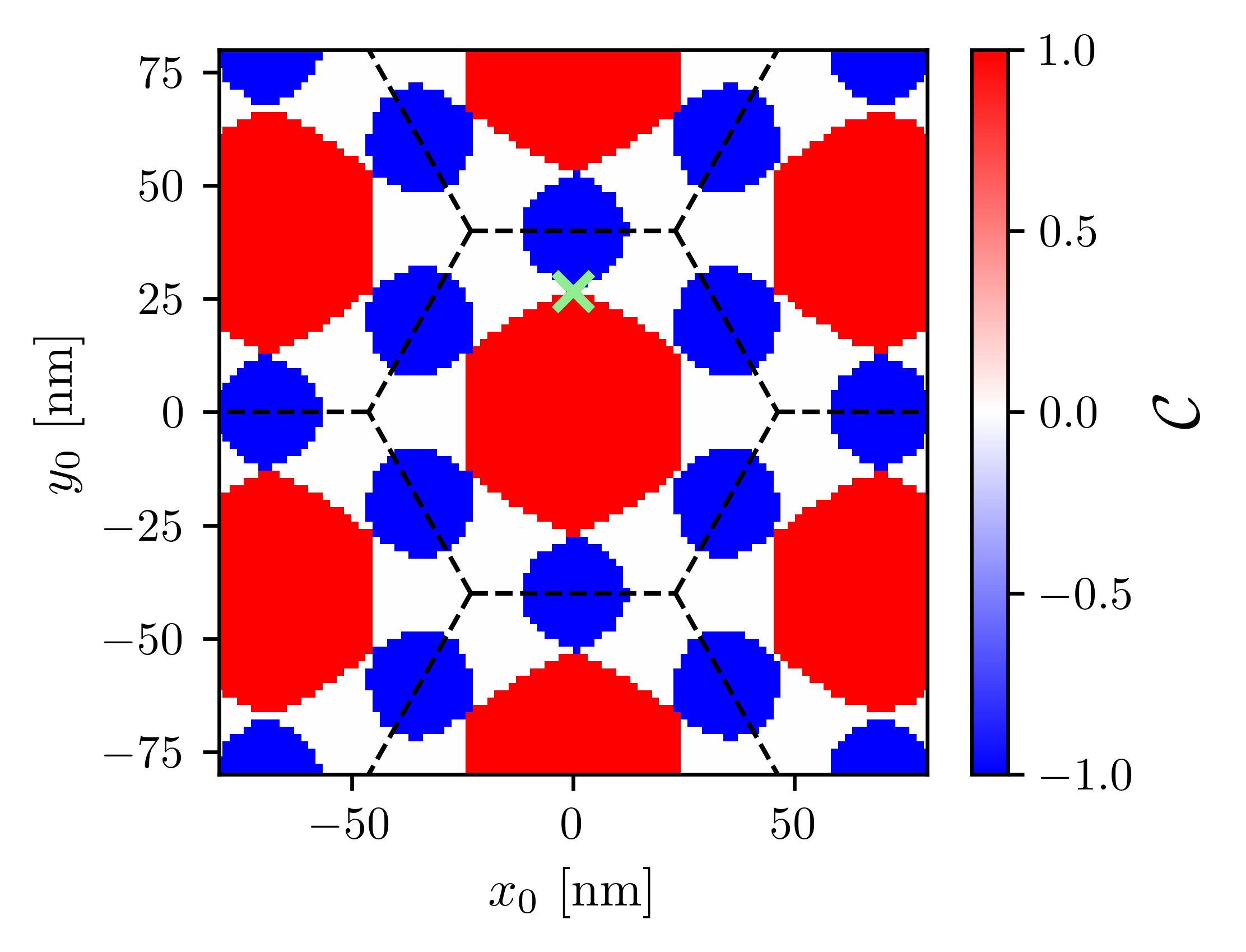}
        \put(0,71){(b)}
    \end{overpic}%
    \caption{Topological phase diagram. The Chern number ${\cal C}$ of the lowest band $\epsilon_{\bm k,\lambda=1}$ as a function of the shift vector $\bm r_0=(x_0,y_0)$. (a) $\beta\Phi_0/(2\pi)=33\ {\rm mT}$. (b) $\beta\Phi_0/(2\pi)=330\ {\rm mT}$. All other parameters are fixed as in Fig.~\ref{fig:Bands}. The light-green cross marks the gap-closing point $\bm r_0^\star$. The dashed black lines outline the hexagonal UC of the direct lattice.}
    \label{fig:phase_diagram}
\end{figure}

Fig.~\ref{fig:phase_diagram} shows the topological phase diagram~\cite{SM}, namely the Chern number ${\cal C}$ of the lowest bands $\epsilon_{\bm k,\lambda =1}$ as a function of the shift vector $\bm r_0 =(x_0,y_0)$ for two values of the magnetic field---$\beta\Phi_0/(2\pi)=33\ {\rm mT}$ in panel (a) and $\beta\Phi_0/(2\pi)=330\ {\rm mT}$ in panel (b).
Red regions correspond to ${\cal C}=+1$, while blue regions to ${\cal C}=-1$.
In both cases, the plots show that the presence of a crystal potential yields
topologically nontrivial bands and a complex but regular phase diagram controlled by $\beta$ and $\bm r_0$.
In particular, for a small value of magnetic field $\beta$, Fig.~\ref{fig:phase_diagram}(a), the system is always topologically non-trivial with either red or blue regions.
On the other hand, for a higher value of $\beta$, Fig.~\ref{fig:phase_diagram}(b), the phase diagram also presents trivial regions (${\cal C}=0$), reported in white.
Note that the red and blue regions are separated by isolated gap-closing points, one of which ($\bm r_0^\star$) is marked as a light-green cross.
We shall see below that an analogy with the Haldane model can be drawn by introducing a low-energy effective model of the Hamiltonian Eq.~\eqref{eq:ham}.

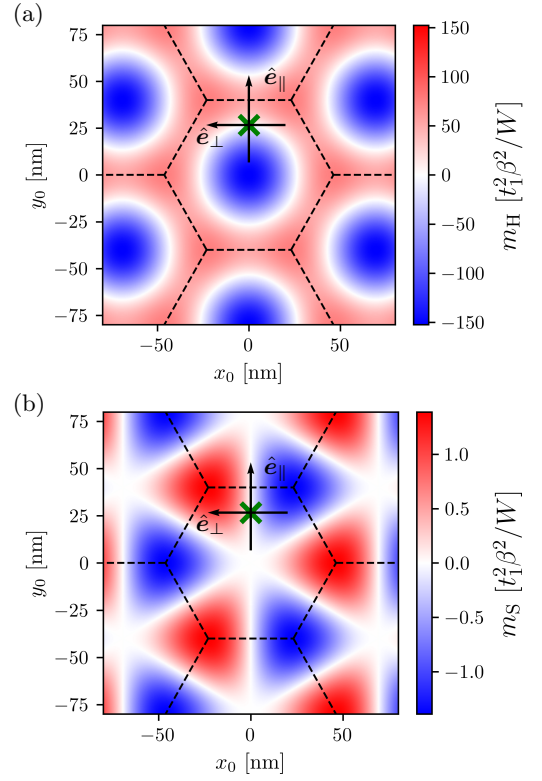
\begin{figure}
    \centering
    \begin{overpic}[width=0.8\columnwidth]{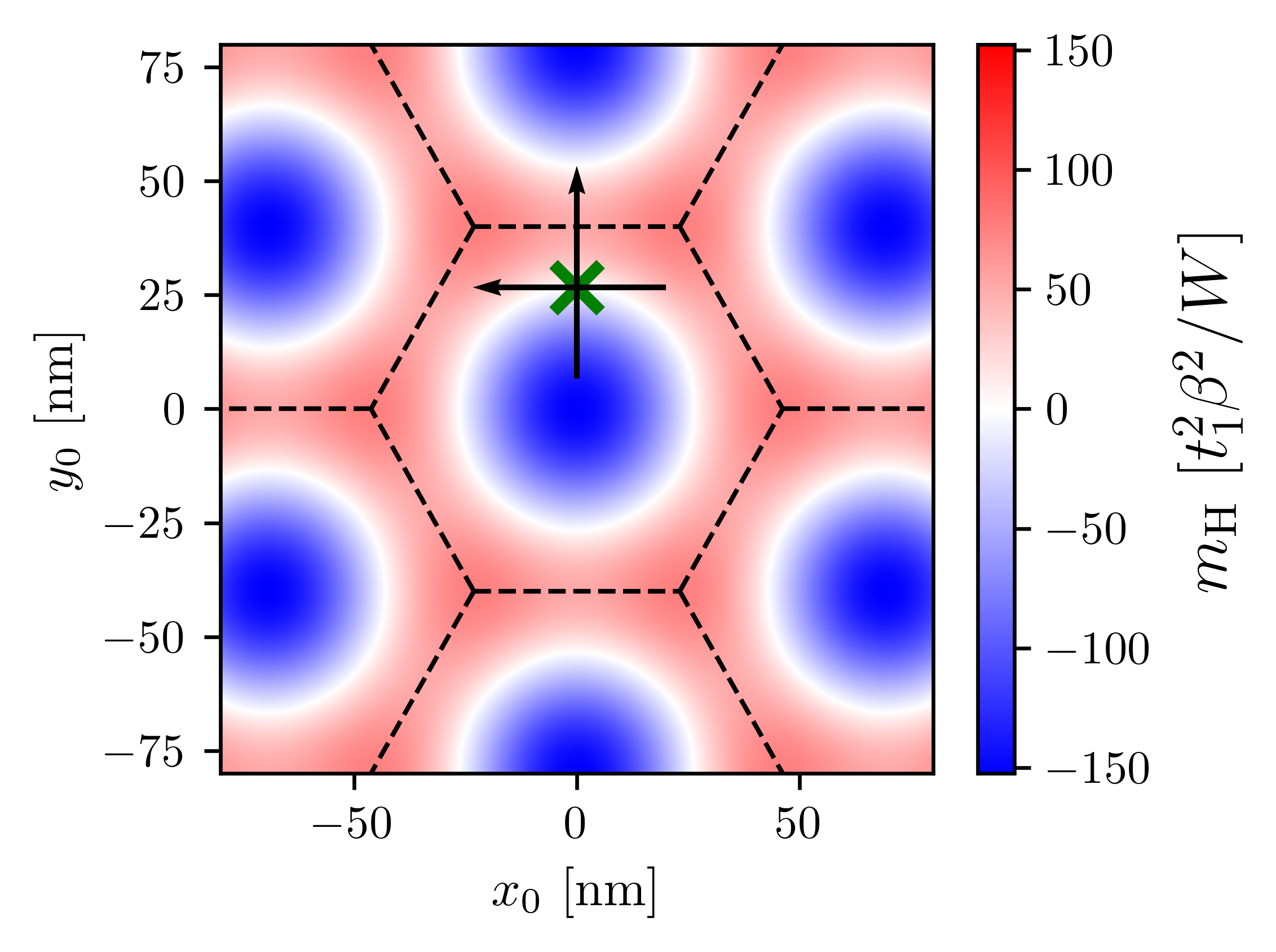}
        \put(0,72){(a)}
        \put(48,60){$\hat{\bm e}_\parallel$}
        \put(35,48.5){$\hat{\bm e}_\perp$}
    \end{overpic}
    \begin{overpic}[width=0.8\columnwidth]{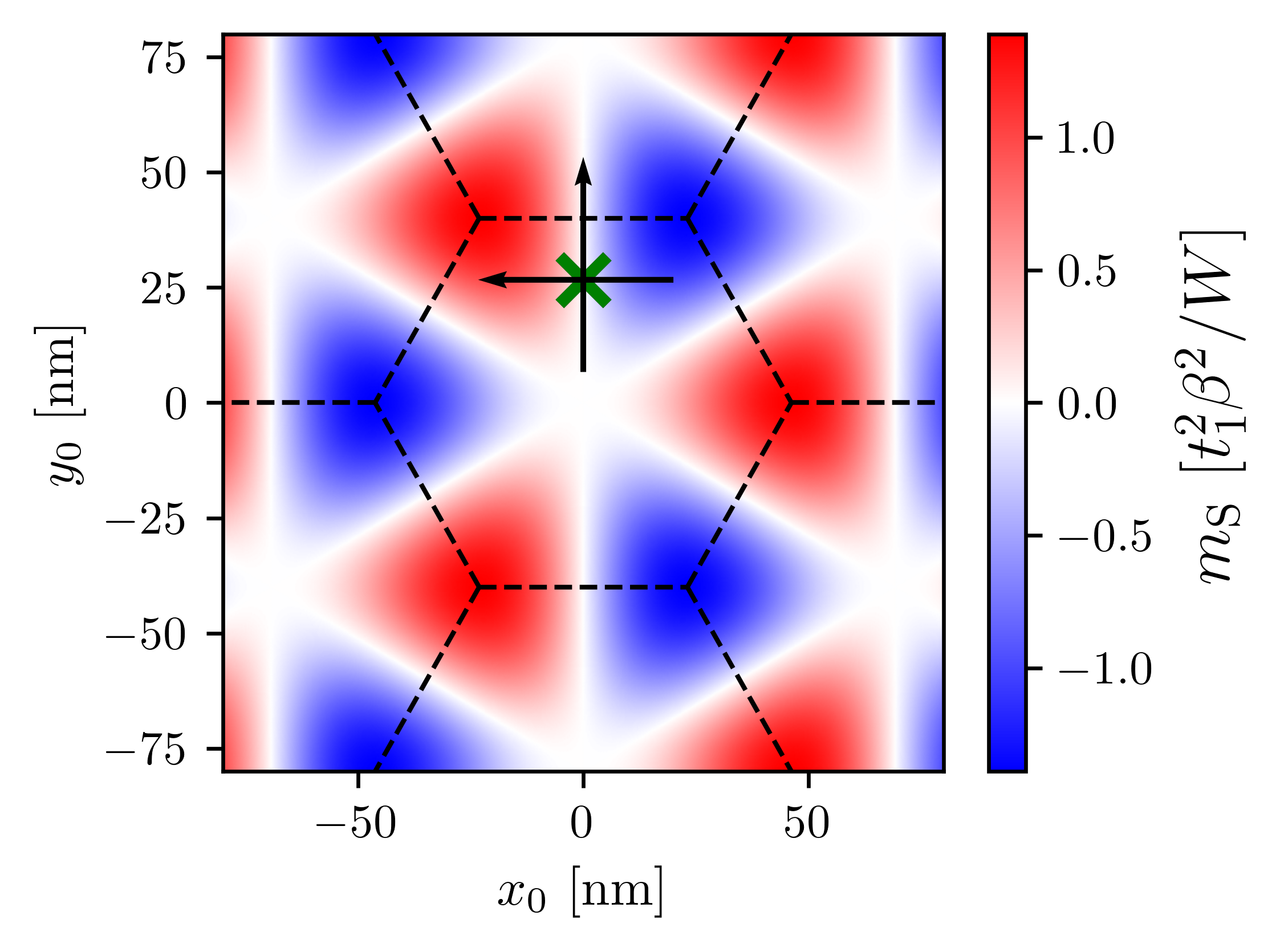}
        \put(0,72){(b)}
        \put(48,60){$\hat{\bm e}_\parallel$}
        \put(35,49){$\hat{\bm e}_\perp$}
    \end{overpic}
    \caption{(a) Haldane mass $m_{\rm H}$ from Eq.~\eqref{eqn:mass_H}. (b) Semenoff mass $m_{\rm S}$ from Eq.~\eqref{eqn:mass_S}. For comparison, we express both masses in units of the energy $t_1^2\beta^2/W$. The black cross marks the gap-closing point $\bm r_0^\star$, and the axes indicate the decoupling directions $\hat{\bm e}_\parallel$ and $\hat{\bm e}_\perp$. Here we fixed $\beta\Phi_0/(2\pi)=33\ {\rm mT}$. All other parameters are fixed as in Fig.~\ref{fig:Bands}.}
    \label{fig:masses}
\end{figure}

{\color{blue}\it Mapping onto the Haldane model.---}
Projecting onto the subspace of Dirac cones near the $K$ and $K'$ valleys, one can approximate Eq.~\eqref{eq:ham} to first order in the momentum deviation from the Dirac points and to second order in the magnetic-field amplitude $\beta$ (see Sect.~\ref{appB} of the SM~\cite{SM}).
The resulting low-energy Hamiltonian is
\begin{equation}
\begin{aligned}
H_{\tau}^{\rm eff}(\bm q)
&= \hbar v\Big[\tau (q_y-q_{0,y})\sigma_x + (q_x-q_{0,x})\sigma_y\Big]\\
&\quad+(\tau m_{\rm H}+m_{\rm S})\,\sigma_z
\\
&\quad+\tau d_0^{(1)}\mathbb{I}
+\mathcal O(q^2,\beta^3)~,
\end{aligned}
\label{eq:Heff_final_with_A2}
\end{equation}
where $\bm q$ denotes a small deviation from $K$ or $K'$, $\tau=\pm 1$ labels the valley, and $\sigma_{i=x,y,z}$ are Pauli matrices acting in the sublattice space. 
Remarkably, and apart from the vector $\bm q_0\equiv(q_{0,x},q_{0,y})$ that originates from the breaking of $C_{3z}$ symmetry induced by a finite $\bm r_0\neq\bm 0$ displacement,
Eq.~\eqref{eq:Heff_final_with_A2} coincides with the low-energy expansion of the Haldane model around $K$ and $K'$ (see Ref.~\cite{Bernevig}).
Such a mapping onto the Haldane model, though, has the peculiarity
that $m_{\rm H}$ (the Haldane mass) and $m_{\rm S}$ (the Semenoff mass) are not independent quantities.
Such masses, in fact, are expressed in terms of the microscopic parameters of Eq.~\eqref{eq:ham} as~\cite{SM}
\begin{subequations}
\begin{equation}
     m_{\rm H} = -\frac{\sqrt{3}t_1}{3}\,\beta\sum_{\ell=1,3,5}\cos(\bm G_\ell\cdot\bm r_0)~,
     \label{eqn:mass_H}
\end{equation}
\begin{equation}
\begin{split}
     m_{\rm S}
     &= \frac{\sqrt{3}}{3}\left[t_{2} + \frac{t_1^2}{9W}\right]\beta^2
     \sum_{\ell=1,3,5}\sin(\bm G_\ell\cdot\bm r_0)
\\
     &\quad-\frac{4\sqrt{3}\,t_1^2}{27W}\,\beta^2
     \prod_{\ell=1,3,5}\sin(\bm G_\ell\cdot\bm r_0)~,
     \label{eqn:mass_S}
\end{split}
\end{equation}
\end{subequations}
where $t_1\equiv \frac{e\hbar}{m^*c}\frac{\Phi_0}{8\pi\sqrt3}$ and
$t_2\equiv\frac{e^2}{8m^*c^2G^2}\left(\frac{\Phi_0}{2\pi}\right)^2$. Both depend non-trivially on $\boldsymbol r_0$ and $\beta$.
Equations~\eqref{eqn:mass_H} and~\eqref{eqn:mass_S} show that while $m_{\rm H}$ is linear in $\beta$ and remains finite in the limit $\bm r_0=\bm 0$, $m_{\rm S}$ is quadratic in $\beta$ and vanishes when $\bm r_0=\bm 0$.
Haldane mass and Semenoff mass are plotted in Figs.~\ref{fig:masses}(a) and (b), respectively, as functions of $x_0$ and $y_0$. 

Note that the Chern number calculated from the Haldane expression ${\cal C} = \frac{1}{2}\left[{\rm sgn}(m_{\rm S}-m_{\rm H})-{\rm sgn}(m_{\rm S}+m_{\rm H})\right]$, using Eqs.~\eqref{eqn:mass_H} and~\eqref{eqn:mass_S}, coincides with the plot in Fig.~\ref{fig:phase_diagram}(a),
obtained with the full model Hamiltonian~(\ref{eq:ham}).


Interestingly, it turns out that the two masses can be tuned independently around some specific points $\bm r_0$.
We first note that both $m_{\rm H}$ and $m_{\rm S}$ vanish at the gap-closing point $\bm r_0^\star = \frac{1}{3}(\bm a_1 + \bm a_2)$, marked by the green cross in Fig.~\ref{fig:masses} and corresponding to the light-green cross shown in Fig.~\ref{fig:phase_diagram}. 
Considering the region in the vicinity of $\bm r_0^\star$, we also note that
%
$m_{\rm H}$ and $m_{\rm S}$ remain zero along orthogonal directions (tangent to the white circle for $m_{\rm H}$ and along the vertical axis for $m_{\rm S}$).
This implies that the two masses can be tuned independently in the neighborhood of $\bm r_0^\star$.
The same behavior holds for the other gap-closing points, obtained from $\bm r_0^\star$ by $C_{6z}$ rotations around the origin.

To make this argument quantitative, we use the approximate expressions \eqref{eqn:mass_H} and \eqref{eqn:mass_S} and expand $m_{\rm H}$ and $m_{\rm S}$ about $\bm r_0^\star$. Let
\begin{equation}
\delta\bm r \equiv \bm r_0-\bm r_0^\star~,
\end{equation}
and define unit vectors that are parallel and perpendicular to $\bm r_0^\star$ (see Fig.~\ref{fig:masses}),
\begin{equation}
\hat{\bm e}_\parallel \equiv \frac{\bm r_0^\star}{\vert \bm r_0^\star\vert },\qquad
\hat{\bm e}_\perp\ \text{such that}\ \hat{\bm e}_\perp\cdot\hat{\bm e}_\parallel=0~,
\end{equation}
so that
\begin{equation}
\delta\bm r=\delta_\parallel\,\hat{\bm e}_\parallel+\delta_\perp\,\hat{\bm e}_\perp~.
\label{eq:delta_r_decomp}
\end{equation}

Expanding the Haldane mass \eqref{eqn:mass_H} in $\delta \bm r$ yields 
\begin{equation}
m_{\rm H}(\bm r_0^\star+\delta\bm r)=
\frac{2\pi t_1}{a_0}\,\beta\,\delta_\parallel
+\mathcal O(\delta_\parallel^2,\delta_\perp^2)~,
\label{eq:m_even_local}
\end{equation}
the leading term being ${\cal O}(\delta_\parallel)$. In particular, notice that there is no linear dependence on $\delta_\perp$. 

Expanding the Semenoff mass~\eqref{eqn:mass_S} up to first order in $\delta \bm r$, one finds 
\begin{equation}
\begin{split}
m_{\rm S}(\bm r_0^\star+\delta\bm r)&=
\frac{2\pi}{a_0}\left(t_{2}+\frac{t_1^2}{9W}\right)\,\beta^2\,\delta_\perp\\
&\quad+\,\mathcal O(\delta_\perp^2,\delta_\parallel\delta_\perp)~,
\label{eq:m_odd_local}
\end{split}
\end{equation}
with no linear dependence on $\delta_\parallel$.
Equations~\eqref{eq:m_even_local} and~\eqref{eq:m_odd_local} show that, in a neighborhood of $\bm r_0^\star$, the Haldane and Semenoff masses are locally decoupled at linear order. Hence, for shifts $\bm r_0$ sufficiently close to $\bm r_0^\star$, the independent control parameters $\delta_\parallel$ and $\delta_\perp$ tune the valley-odd and valley-even mass components separately at leading order.
Furthermore, consider a vertical cut through the point $\bm r_0^\star = \frac{1}{3}(\bm a_1 + \bm a_2)$ in the phase diagram of Fig.~\ref{fig:phase_diagram}(b). The resulting profile closely resembles the Haldane phase diagram at $\varphi=\pi$~\cite{Haldane1988}, which separates the ${\cal C}=-1$ phase from the ${\cal C}=1$ phase.
Remarkably, this shows that even for arbitrary values of the PMFs, we find a direct analogy with the Haldane model by focusing on regions of the phase diagram around the gap-closing points.




{\color{blue} \it Summary and conclusions.---}We studied a minimal model of spinless, non-interacting electrons in two spatial dimensions subject to a periodic scalar potential $V(\bm r)$ and a PMF $B_z(\bm r)$ with zero net flux per UC. The vanishing-flux condition allows for a periodic gauge and preserves ordinary lattice translations for any relative displacement $\bm r_0$ between the two patterns $V(\bm r)$ and $B_z(\bm r)$, providing topological Bloch bands without a uniform magnetic field.

Numerical plane-wave diagonalization shows that the PMF opens gaps at the Dirac points of the underlying honeycomb-like low-energy structure and generates extended regions of nonzero Chern number. The resulting topological phase diagram is controlled by the field amplitude $\beta$ and by the shift $\bm r_0$.


A low-energy projection near $K$ and $K'$ maps the model onto a Haldane-type massive Dirac theory with a valley-odd Haldane mass $m_{\rm H}$ and a valley-even Semenoff mass $m_{\rm S}$. Microscopically, $m_{\rm H}$ appears at $\mathcal O(\beta)$, whereas $m_{\rm S}$ arises at $\mathcal O(\beta^2)$ and vanishes as $\bm r_0\to\bm 0$, explaining the weak-field phase boundaries dominated by $m_{\rm H}=0$ and the stronger-field topological properties induced by $m_{\rm S}$. Although the two masses are generally linked functions of $(\beta,\bm r_0)$, near special gap-closing points $\bm r_0^\star$, they decouple to linear order, such that orthogonal displacements independently tune $m_{\rm H}$ and $m_{\rm S}$, recovering the canonical Haldane control parameters locally.

These results identify the misalignment $\bm r_0$ between the periodic scalar potential $V(\bm r)$ and the PMF $B_z(\bm r)$ as a purely geometric knob to switch between Chern phases and to engineer chiral edge physics in experimentally relevant parameter ranges.
Concerning possible experimental implementations, we stress that the choice of PMF considered here is only one example leading to
topologically non trivial phases.
Indeed, as shown in the SM~\cite{SM}, a different choice of PMF yields a phase diagram with extended regions of nonzero Chern number.
Extensions to include spin, interactions, and disorder will clarify the robustness of our results and are left for future work.

{\it \color{blue} Acknowledgments.---}It is a pleasure to thank Oleg Sushkov and Alex Hamilton for useful discussions. M.P. was supported by the European Union under grant agreement No. 101131579 - Exqiral. Views and opinions expressed are however those of the
author(s) only and do not necessarily reflect those of the European Union or the European Commission. Neither the European Union nor the granting authority can
be held responsible for them. F.T. acknowledges funding from MUR-PRIN 2022 - Grant No. 2022B9P8LN - (PE3)- Project NEThEQS ``Non-equilibrium coherent thermal effects in quantum systems” in PNRR Mission 4 - Component 2 - Investment 1.1 “Fondo per il Programma Nazionale di Ricerca e Progetti di Rilevante Interesse Nazionale (PRIN)'' funded by the European Union - Next Generation EU and the PNRR MUR project PE0000023-NQSTI, and from the Royal Society through the International Exchanges between the UK and Italy (Grant No. IEC R2 192166).

\clearpage
\onecolumngrid            
\setcounter{page}{1}

\setcounter{secnumdepth}{3} 
\setcounter{section}{0}
\setcounter{subsection}{0}
\setcounter{subsubsection}{0}
\setcounter{equation}{0}
\setcounter{figure}{0}
\setcounter{table}{0}

\makeatletter
\renewcommand{\thesection}{S\arabic{section}}
\renewcommand{\thesubsection}{S\arabic{section}.\arabic{subsection}}
\renewcommand{\thesubsubsection}{S\arabic{section}.\arabic{subsection}.\arabic{subsubsection}}

\renewcommand{\theequation}{S\arabic{section}.\arabic{equation}}
\renewcommand{\thefigure}{S\arabic{figure}}
\renewcommand{\thetable}{S\arabic{table}}

\@ifundefined{theHsection}{}{%
  \renewcommand{\theHsection}{S\arabic{section}}
  \renewcommand{\theHsubsection}{S\arabic{section}.\arabic{subsection}}
  \renewcommand{\theHsubsubsection}{S\arabic{section}.\arabic{subsection}.\arabic{subsubsection}}
  \renewcommand{\theHequation}{S\arabic{equation}}
  \renewcommand{\theHfigure}{S\arabic{figure}}
  \renewcommand{\theHtable}{S\arabic{table}}
}

\def\@seccntformat#1{\csname the#1\endcsname\quad}
\makeatother

\renewcommand{\bibnumfmt}[1]{[S#1]}
\renewcommand{\citenumfont}[1]{S#1}

\begin{center}
\textbf{\Large Supplemental Material for:\\ ``Simulating the Haldane model on ultra-clean GaAs heterostructures''}

\bigskip

Francesco Cioni,$^{1}$
Lorenzo Cavicchi,$^{2}$
Fabio Taddei,$^{3}$
Marco Polini$^{4,\,5}$

\bigskip

$^1$\,{\it NEST, Scuola Normale Superiore, I-56126 Pisa,~Italy}

$^2$\,{\it Scuola Normale Superiore, Piazza dei Cavalieri 7, I-56126 Pisa,~Italy}

$^3$\,{\it NEST, Istituto Nanoscienze-CNR, Piazza S. Silvestro 12, I-56126 Pisa,~Italy}

$^4$\,{\it Dipartimento di Fisica dell'Universit\`a di Pisa, Largo Bruno Pontecorvo 3, I-56127 Pisa,~Italy}

$^5$\,{\it ICFO-Institut de Ci\`{e}ncies Fot\`{o}niques, The Barcelona Institute of Science and Technology, Av. Carl Friedrich Gauss 3, 08860 Castelldefels (Barcelona),~Spain}

%
%

%

\bigskip

In this Supplemental Material we present details on the plane-wave expansion method used to diagonalize the Hamiltonian of Eq.~\eqref{eq:ham}, the calculation of the Chern number, the detailed $K$-point three-plane-wave expansion used to derive the low-energy Hamiltonian of Eq.~\eqref{eq:Heff_final_with_A2}, and results concerning the topological phase diagram for different periodic magnetic field patterns.
 
\end{center}

\onecolumngrid

\section{Plane-wave expansion}\label{appA}
In order to diagonalize the Hamiltonian in Eq.~\eqref{eq:ham}, we perform a standard plane-wave expansion. In the following, we calculate the matrix elements of the Hamiltonian between two generic plane waves defined by the wavevectors $\bm{k}$ and $\bm{k}'$.
To calculate such matrix elements it is more useful to write the crystal and vector potential as:

\begin{eqnarray}\label{eq: V_exp}
    V(\bm{r}) &=& W \sum_{\ell=1,3,5} \left(e^{i\bm{G}_\ell \cdot \bm{r} } + e^{-i\bm{G}_\ell \cdot \bm{r}}\right) ~ , \\ \label{eq: A_exp}
    \bm{A}(\bm{r}-\bm r_0) &=& \frac{\beta \Phi_0}{2\pi} \sum_{\ell=1,3,5} \frac{\bm{G}_\ell \wedge \hat{\bm{z}}}{2i \vert  \bm{G}_\ell \vert ^2} \Big ( e^{i\bm{G}_\ell \cdot (\bm{r} - \bm r_0)} - e^{-i\bm{G}_\ell \cdot (\bm{r} - \bm r_0)} \Big ) \ .
\end{eqnarray}
It is now easy to calculate matrix elements between plane waves using the identity:
\begin{eqnarray}
    \langle \bm{k} \vert  e^{i\bm{q} \cdot \hat{\bm{r}}} \vert  \bm{k}' \rangle = \frac{1}{A_{\rm u.c.}} \int d^2 r \ e^{i(-\bm{k} + \bm{q} + \bm{k}') \cdot \bm{r}} = \delta_{\bm{k}', \bm{k}-\bm{q}} \ .
\end{eqnarray}

As a first step, we expand the square in the Hamiltonian in Eq. \eqref{eq:ham}:
\begin{equation}\label{eqn:ham_SM}
    \hat{H} = \frac{\hat{p}^2}{2m^*} + \frac{e}{2m^*c}\Big ( \hat{\bm{p}} \cdot \bm{A}(\hat{\bm{r}}-\bm r_0) + \bm{A}(\hat{\bm{r}}-\bm r_0) \cdot \hat{\bm{p}} \Big ) + \frac{e^2}{2m^*c^2} A^2(\hat{\bm{r}}-\bm r_0) + V(\hat{\bm{r}}) \ .
\end{equation}
The first term provides only diagonal contributions:
\begin{equation}
    \langle \bm{k} \vert  \frac{\hat{p}^2} {2m^*} \vert  \bm{k}' \rangle = \frac{\hbar^2 k^2}{2m^*} \delta_{\bm{k}, \bm{k}'} \ .
\end{equation}
The second one gives:
\begin{eqnarray}
    &&\frac{e}{2m^*c}\langle \bm{k} \vert  \hat{\bm{p}} \cdot \bm{A}(\hat{\bm{r}}-\bm r_0) + \bm{A}(\hat{\bm{r}}-\bm r_0) \cdot \hat{\bm{p}} \vert  \bm{k}' \rangle = \frac{e\hbar}{2m^*c}(\bm{k} + \bm{k}') \cdot \langle \bm{k} \vert  \bm{A}(\hat{\bm{r}}) \vert  \bm{k}' \rangle = \\
    &=& \frac{e\hbar}{2m^*c}\frac{\beta \Phi_0}{2\pi} \sum_{\ell=1,3,5} (\bm{k} + \bm{k}') \cdot\frac{(\bm{G}_\ell \wedge \hat{\bm{z}})}{2i \vert  \bm{G}_\ell \vert ^2} \Big ( e^{-i\bm{G}_\ell \cdot \bm r_0} \delta_{\bm{k}, \bm{G}_\ell + \bm{k}'} -e^{i\bm{G}_\ell \cdot \bm r_0} \delta_{\bm{k}, -\bm{G}_\ell + \bm{k}'}  \Big ) = \\
    &=& \frac{e\hbar}{2m^*c}\frac{\beta \Phi_0}{2\pi} \sum_{\ell=1,3,5} \Big [ (2\bm{k}' + \bm{G}_\ell) \cdot\frac{(\bm{G}_\ell \wedge \hat{\bm{z}})}{2i \vert  \bm{G}_\ell \vert ^2}  e^{-i\bm{G}_\ell \cdot \bm r_0}\delta_{\bm{k}, \bm{G}_\ell + \bm{k}'} -(2\bm{k}' - \bm{G}_\ell) \cdot\frac{(\bm{G}_\ell \wedge \hat{\bm{z}})}{2i \vert  \bm{G}_\ell \vert ^2}  e^{i\bm{G}_\ell \cdot \bm r_0}\delta_{\bm{k}, -\bm{G}_\ell + \bm{k}'}  \Big ]  = \\
    &=& \frac{-ie\hbar}{2m^*c}\frac{\beta \Phi_0}{2\pi} \sum_{\ell=1,3,5} \bm{k}' \cdot\frac{(\bm{G}_\ell \wedge \hat{\bm{z}})}{\vert  \bm{G}_\ell \vert ^2}  \Big ( e^{-i\bm{G}_\ell \cdot \bm r_0}\delta_{\bm{k}- \bm{G}_\ell, \bm{k}'} - e^{i\bm{G}_\ell \cdot \bm r_0} \delta_{\bm{k}+ \bm{G}_\ell, \bm{k}'}  \Big ) \ , 
\end{eqnarray}
where Eq. \eqref{eq: A_exp} is used in passing from the first to the second line.
Similarly, using Eq. \eqref{eq: V_exp}, the matrix elements of the crystal potential are given by:
\begin{eqnarray}
    \langle \bm{k} \vert  V(\hat{\bm{r}}) \vert  \bm{k}' \rangle = W \sum_{\ell=1,3,5} \Big (\delta_{\bm{k}-\bm{G}_\ell, \bm{k}'} +\delta_{\bm{k}+\bm{G}_\ell, \bm{k}'}  \Big ) \ .
\end{eqnarray}
Finally, the $A^2$ term in Eq.~(\ref{eqn:ham_SM}) is a bit more involved due to the fact that two reciprocal vectors are exchanged in the process. Indeed, we have:
\begin{eqnarray}
    &&A^2(\bm{r}-\bm r_0) = \\
    &=& \Big (\frac{\beta \Phi_0}{2\pi} \Big )^2 \sum_{\ell,m=1,3,5} \frac{\bm{G}_\ell \wedge \hat{\bm{z}}}{2i \vert  \bm{G}_\ell \vert ^2} \cdot\frac{\bm{G}_m \wedge \hat{\bm{z}}}{2i \vert  \bm{G}_m \vert ^2} \Big ( e^{i(\bm{G}_\ell + \bm{G}_m) \cdot (\bm{r} -\bm r_0)} - e^{i(\bm{G}_\ell - \bm{G}_m) \cdot (\bm{r} - \bm r_0)} - e^{-i(\bm{G}_\ell - \bm{G}_m) \cdot (\bm{r} - \bm r_0)} + e^{-i(\bm{G}_\ell + \bm{G}_m)\cdot (\bm{r} - \bm r_0)} \Big ) \nonumber\\
    &=& -\frac{1}{4\vert  \bm{G} \vert ^4} \Big (\frac{\beta \Phi_0}{2\pi} \Big )^2 \sum_{\ell,m=1,3,5} (\bm{G}_\ell \cdot \bm{G}_m)\Big ( e^{i(\bm{G}_\ell + \bm{G}_m) \cdot (\bm{r} - \bm r_0)} - e^{i(\bm{G}_\ell - \bm{G}_m) \cdot (\bm{r} - \bm r_0)} - e^{-i(\bm{G}_\ell - \bm{G}_m) \cdot (\bm{r} - \bm r_0)} + e^{-i(\bm{G}_\ell + \bm{G}_m)\cdot (\bm{r} - \bm r_0)} \Big ) \nonumber\ ,
\end{eqnarray}
and the matrix elements between plane waves follow directly:
\begin{equation}
\begin{split}
   \frac{e^2}{2m^*c^2}&\langle \bm{k} \vert  A^2(\hat{\bm{r}}) \vert  \bm{k}' \rangle = -\frac{e^2}{2m^*c^2}\frac{1}{4\vert  \bm{G} \vert ^2} \Big (\frac{\beta \Phi_0}{2\pi} \Big )^2 \sum_{\ell,m=1,3,5} \frac{\bm{G}_\ell \cdot \bm{G}_m}{\vert  \bm{G} \vert ^2}\Big ( e^{-i(\bm{G}_\ell + \bm{G}_m) \cdot \bm r_0}\delta_{\bm{k}, \bm{G}_\ell + \bm{G}_m + \bm{k}'} \\
   &- e^{-i(\bm{G}_\ell - \bm{G}_m) \cdot \bm r_0}\delta_{\bm{k}, \bm{G}_\ell - \bm{G}_m + \bm{k}'} - e^{i(\bm{G}_\ell - \bm{G}_m) \cdot \bm r_0}\delta_{\bm{k}, -\bm{G}_\ell + \bm{G}_m + \bm{k}'} + e^{i(\bm{G}_\ell + \bm{G}_m) \cdot \bm r_0}\delta_{\bm{k}, -\bm{G}_\ell - \bm{G}_m + \bm{k}'} \Big ) \ .
\end{split}
\end{equation}

\section{Chern number calculation}
In this Section, we provide details on the numerical calculation of the Chern number ${\cal C}$. In order to obtain the results for the topological phase diagram in Fig.~\ref{fig:phase_diagram} of the main text, we employ the Fukui-Hatsugai-Suzuki (FHS) algorithm~\cite{fukui2005}, which we briefly summarize in the following.

Consider a 2D periodic Bloch Hamiltonian $H({\bm k})$ with ${\bm k}=(k_x,k_y)$ in the Brillouin zone (BZ).
Let $\{\vert  u_n({\bm k})\rangle\}$ be the cell-periodic eigenstates of an isolated band (or an isolated group of bands).
The first Chern number of that band is, in the continuum,
\begin{equation}
{\cal C}=\frac{1}{2\pi}\int_{\mathrm{BZ}} d^2k\; \Omega_{xy}({\bm k}),
\qquad
\Omega_{xy}=\partial_{k_x}\mathcal{A}_y-\partial_{k_y}\mathcal{A}_x,
\qquad
\mathcal{A}_\mu=i\langle u\vert \partial_{k_\mu}u\rangle,
\end{equation}
which is gauge-invariant but numerically delicate because it involves first derivatives of the eigenstates and depends on the phase-gauge choice.
The FHS algorithm evaluates ${\cal C}$ on a discrete ${\bm k}$-mesh in an explicit gauge-invariant way. First, one discretizes the BZ into a uniform grid ${\bm k}_{m,n}$ with steps $\Delta k_x,\Delta k_y$.
Then, one defines the overlap variables for a single, isolated band as
\begin{equation}
U_x({\bm k})=\frac{\langle u({\bm k})\vert  u({\bm k}+\Delta k_x\hat{x})\rangle}
{\big\vert \langle u({\bm k})\vert  u({\bm k}+\Delta k_x\hat{x})\rangle\big\vert },
\qquad
U_y({\bm k})=\frac{\langle u({\bm k})\vert  u({\bm k}+\Delta k_y\hat{y})\rangle}
{\big\vert \langle u({\bm k})\vert  u({\bm k}+\Delta k_y\hat{y})\rangle\big\vert }.
\end{equation}
Each $U_\alpha({\bm k})$ is the phase of the overlap between neighboring Bloch states. Therefore, it is well-defined
even if $\vert  u({\bm k})\rangle$ has an arbitrary ${\bm k}$-dependent phase.
On each elementary plaquette with a corner at ${\bm k}$, define the discrete Berry curvature
\begin{equation}
\tilde{\Omega}_{xy}({\bm k})
=\mathrm{Arg}\!\left[
U_x({\bm k})\,U_y({\bm k}+\Delta k_x\hat{x})\,
U_x({\bm k}+\Delta k_y\hat{y})^{*}\,U_y({\bm k})^{*}
\right],
\label{eq:FHScurv}
\end{equation}
where $\mathrm{Arg}(z)\in(-\pi,\pi]$ denotes the principal branch of the complex argument.
The product inside $\mathrm{Arg}$ is such that the gauge phases cancel exactly, so
$\tilde{\Omega}_{xy}({\bm k})$ is strictly gauge-invariant on the grid.
The FHS Chern number is obtained by summing the lattice field strength over all plaquettes:
\begin{equation}
C_{\mathrm{FHS}}=\frac{1}{2\pi}\sum_{{\bm k}\in {\rm grid}}\tilde{\Omega}_{xy}({\bm k}).
\label{eq:FHSchern}
\end{equation}
With periodic boundary conditions on the mesh, $C_{\mathrm{FHS}}$ is an integer, provided that the band remains isolated.
Taking the continuum limit, $\tilde{\Omega}_{xy}({\bm k})\approx \Omega_{xy}({\bm k})\,d k_xd k_y$
and $C_{\mathrm{FHS}}\to {\cal C}$.

\section{Expansion around the $K$-point}\label{appB}
In this Section, we provide details on the full derivation of the continuum Hamiltonian as obtained by expanding around the $K$ and $K^\prime$ valleys.

Our starting point is the full Hamiltonian in Eq.~\eqref{eqn:ham_SM}.
Consider the reciprocal vectors
\begin{eqnarray}
\bm G_1&=&\Big(\frac{2\pi}{\sqrt3 a_0},\,\frac{2\pi}{a_0}\Big),\\
\bm G_2&=&\Big(-\frac{2\pi}{\sqrt3 a_0},\,\frac{2\pi}{a_0}\Big) = - \bm G_5,\\
-\bm G_3&\equiv& \bm G_1-\bm G_2=\Big(\frac{4\pi}{\sqrt3 a_0},\,0\Big),
\end{eqnarray}
with $\vert \bm G_1\vert =\vert \bm G_2\vert =\vert \bm G_3\vert \equiv G=4\pi/(\sqrt3 a_0)$.
The BZ corners can be taken as
\begin{equation}
\bm K=\frac{\bm G_1+\bm G_2}{3}=\Big(0,\,\frac{4\pi}{3a_0}\Big),\qquad \bm K'=-\bm K.
\end{equation}
The three plane waves
\begin{equation}
\vert 1\rangle\equiv\vert \bm K\rangle,\qquad \vert 2\rangle\equiv\vert \bm K-\bm G_1\rangle,\qquad \vert 3\rangle\equiv\vert \bm K-\bm G_2\rangle
\label{eq:three_states}
\end{equation}
have equal kinetic energy
\begin{equation}
E_0\equiv \frac{\hbar^2\vert  {\bm K}\vert ^2}{2m^*},\qquad \vert \bm K\vert ^2=\vert \bm K-\bm G_1\vert ^2=\vert \bm K-\bm G_2\vert ^2.
\end{equation}
Moreover, the crystal potential couples them with an amplitude $W$ because
\begin{eqnarray}
(\bm K)-(\bm K-\bm G_1)&=&\bm G_1,\\
(\bm K)-(\bm K-\bm G_2)&=&\bm G_2,\\
(\bm K-\bm G_1)-(\bm K-\bm G_2)&=&-(\bm G_1-\bm G_2)=\bm G_3,
\end{eqnarray}

In the basis \eqref{eq:three_states}, the projected crystal potential is expressed as the $3\times 3$ matrix 
\begin{equation}
H_V=W\begin{pmatrix}
0&1&1\\
1&0&1\\
1&1&0
\end{pmatrix}.
\label{eq:HV3}
\end{equation}
Diagonalizing $H_V$ in Eq.~\eqref{eq:HV3} gives the following eigenstates:
\begin{align}
\vert s\rangle &=\frac{1}{\sqrt3}(1,1,1)^T,\qquad &&E_s=+2W,\\
\vert u_+\rangle &=\frac{1}{\sqrt3}(1,\omega,\omega^2)^T,\qquad &&E_d=-W,\\
\vert u_-\rangle &=\frac{1}{\sqrt3}(1,\omega^2,\omega)^T,\qquad &&E_d=-W,
\end{align}
where $\omega=e^{i2\pi/3}$.
Thus, without the magnetic vector potential $\bm A$, the low-energy subspace at $K$ is the degenerate doublet $\{\vert u_+\rangle,\vert u_-\rangle\}$ at energy $E_0-W$.

In what follows, we restrict ourselves to the Hilbert space spanned by the three plane-waves in Eq.~\eqref{eq:three_states} and treat the PMF using perturbation theory. We derive the continuum approximation of the full Hamiltonian~\eqref{eq:ham} inside the doublet subspace to capture the Dirac cone physics. Henceforth, we denote by $\hat P$ the projector onto this doublet and by $\hat Q=\hat \openone - \hat P$ its complementary. 
To proceed, we divide the Hamiltonian~\eqref{eq:ham} into contributions of first and second order in $\bm A$:
\begin{align}
\hat{\mathcal H}
&=\frac{\hat{\bm p}^2}{2m^*}+V(\hat{\bm r})+\hat H_A^{(1)}+\hat H_A^{(2)},\\
\hat H_A^{(1)}&=\frac{e}{2m^*c}\,(\bm A(\hat{\bm r}-\bm r_0)\cdot \hat{\bm p} + \hat{\bm p}\cdot \bm A(\hat{\bm r}-\bm r_0)),\\
\hat H_A^{(2)}&=\frac{e^2}{2m^*c^2}\,A^2(\hat{\bm r}-\bm r_0).
\end{align}
In the following sub-sections, we will obtain results to first order, incorporating the effects of the paramagnetic term $\hat H_A^{(1)}$, and then include the leading $\mathcal O(A^2)$ effect on the low-energy doublet through a Schrieffer-Wolff (SW) transformation, finally adding the diamagnetic contribution $\hat H_A^{(2)}$.

\subsection{Derivation of the Dirac Hamiltonian}\label{appB1}
We set $\bm k=\bm K+\bm q$ with $\vert \bm q\vert \ll \vert \bm K\vert $ and define
\begin{equation}
\bm k_1\equiv \bm K,\qquad \bm k_2\equiv \bm K-\bm G_1,\qquad \bm k_3\equiv \bm K-\bm G_2.
\end{equation}
Then, introducing $K\equiv\vert \bm K\vert =\frac{4\pi}{3a_0}$, one finds the following convenient coordinate form
\begin{eqnarray}
\bm k_1&=&(0,\,K),\\
\bm k_2&=&\Big(-\frac{\sqrt3}{2}K,\,-\frac12 K\Big),\\
\bm k_3&=&\Big(+\frac{\sqrt3}{2}K,\,-\frac12 K\Big),
\end{eqnarray}
i.e. the three vectors are related by $2\pi/3$ rotations and satisfy $\vert \bm k_1\vert =\vert \bm k_2\vert =\vert \bm k_3\vert =K$.
For each plane wave $\vert \bm k_j+\bm q\rangle$, we have
\begin{equation}
\begin{split}
\frac{\hbar^2}{2m^*}\vert \bm k_j+\bm q\vert ^2
&=\frac{\hbar^2}{2m^*}\left(K^2+2\,\bm k_j\cdot \bm q+\vert \bm q\vert ^2\right)\\
&=E_0+\delta E_j(\bm q)+\mathcal O(q^2),
\end{split}
\end{equation}
where $E_0=\hbar^2K^2/(2m^*)$ and
\begin{equation}
\delta E_j(\bm q)=\frac{\hbar^2}{m^*}\,\bm k_j\cdot \bm q.
\end{equation}
Thus, within the three-state manifold, the $\bm k\cdot \bm p$ correction from $\hat{\bm p}^2/(2m^*)$ is the diagonal matrix
\begin{equation}
\delta H_{\rm kin}(\bm q)=\delta E_i\delta_{ij}.
\end{equation}
Using the explicit $\bm k_j$ above and $\bm q=(q_x,q_y)$, we have
\begin{align*}
\delta E_1 &= \frac{\hbar^2}{m^*}\,K\,q_y,\\
\delta E_2 &= \frac{\hbar^2}{m^*}\,K\left(-\frac{\sqrt3}{2}q_x-\frac12 q_y\right),\\
\delta E_3 &= \frac{\hbar^2}{m^*}\,K\left(+\frac{\sqrt3}{2}q_x-\frac12 q_y\right).
\end{align*}

As we have seen, diagonalizing the scalar potential $H_V$ yields the singlet $\vert s\rangle$ and the doublet
\begin{equation}
\begin{split}
\vert  u_+\rangle=\frac{1}{\sqrt3}(1,\omega,\omega^2)^{\rm T},\qquad
\vert  u_-\rangle=\frac{1}{\sqrt 3}(1,\omega^2,\omega)^{\rm T}.
\end{split}
\end{equation}
The projector on the degenerate subspace is explicitly $\hat P=\vert u_+\rangle\langle u_+\vert +\vert u_-\rangle\langle u_-\vert $.
The effective kinetic term in the doublet is the $2\times 2$ matrix
\begin{equation}
\begin{split}
H_{\rm kin}^{\rm (dbl)}(\bm q)&=\hat P\,\delta H_{\rm kin}(\bm q)\,\hat P
\\
&\equiv
\begin{pmatrix}
\langle u_+\vert \delta H_{\rm kin}\vert u_+\rangle & \langle u_+\vert \delta H_{\rm kin}\vert u_-\rangle\\
\langle u_-\vert \delta H_{\rm kin}\vert u_+\rangle & \langle u_-\vert \delta H_{\rm kin}\vert u_-\rangle
\end{pmatrix}.
\end{split}
\end{equation}
Since $\delta H_{\rm kin}$ is diagonal,
\begin{equation}
\langle u_\mu\vert \delta H_{\rm kin}\vert u_\nu\rangle
=\sum_{j=1}^3 (u_\mu)_j^*\,\delta E_j\,(u_\nu)_j.
\end{equation}

Because $\vert (u_\pm)_j\vert ^2=1/3$ for all $j$,
\begin{equation}
\langle u_\pm\vert \delta H_{\rm kin}\vert u_\pm\rangle
=\frac{1}{3}(\delta E_1+\delta E_2+\delta E_3).
\end{equation}
Using the explicit $\delta E_j$ above, we have
\begin{equation}
\delta E_1+\delta E_2+\delta E_3
=\frac{\hbar^2K}{m^*}\left[q_y-\frac12 q_y-\frac12 q_y\right]
+\frac{\hbar^2K}{m^*}\left[-\frac{\sqrt3}{2}q_x+\frac{\sqrt3}{2}q_x\right]=0,
\end{equation}
hence
\begin{equation}
\langle u_\pm\vert \delta H_{\rm kin}\vert u_\pm\rangle=0.
\end{equation}
Using $(u_+)_2^* (u_-)_2=\omega$ and $(u_+)_3^*(u_-)_3=\omega^{-1}=\omega^{-1}e^{2\pi i}=\omega^2$, we have
\begin{equation}
\langle u_+\vert \delta H_{\rm kin}\vert u_-\rangle
=\frac13\left(\delta E_1+\omega\,\delta E_2+\omega^2\,\delta E_3\right).
\end{equation}
Substituting the $\delta E_j$ and $\omega=-\frac12+i\frac{\sqrt3}{2}$ gives the closed form
\begin{gather}
\langle u_+\vert \delta H_{\rm kin}\vert u_-\rangle
=\frac{\hbar^2K}{2m^*}\,(q_y-i q_x),
\\
\langle u_-\vert \delta H_{\rm kin}\vert u_+\rangle
=\frac{\hbar^2K}{2m^*}\,(q_y+i q_x).
\end{gather}
Therefore

\begin{equation}
    H_{\rm kin}^{\rm (dbl)}(\bm q)=\hbar v\,(q_y\sigma_x+q_x\sigma_y),
\end{equation}
with $v=\frac{\hbar K}{2m^*}=\frac{2\pi\hbar}{3a_0 m^*}$.
Adding the unperturbed doublet energy $(E_0-W)$, the effective Hamiltonian at valley $K$, to linear order in $\bm q$ and before including the magnetic vector potential $\bm A$, is
\begin{equation}\label{eq:Dirac_kin}
H^{(0)}_{K}(\bm q)=(E_0-W)\openone+\hbar v\,(q_y\sigma_x+q_x\sigma_y)+\mathcal O(q^2)\ , 
\end{equation}
where the zero in $H^{(0)}_{K}$ corresponds to the order of the expansion in the external magnetic field.
A different choice of basis within the doublet (corresponding to a unitary rotation) can bring this into the more conventional form
$\hbar v(q_x\sigma_x+q_y\sigma_y)$. The physics---Dirac cone with velocity $v$---is unchanged.

\subsection{Magnetic perturbation at $K$: first order perturbation in $A$}
Define the three shift phases
\begin{eqnarray}
\phi_1&\equiv& \bm G_1\cdot \bm r_0,\\
\phi_2&\equiv& \bm G_2\cdot \bm r_0,\\
\phi_3&\equiv& -\bm G_3\cdot \bm r_0=\phi_1-\phi_2.
\label{eq:phis}
\end{eqnarray}
Using Eq.~\eqref{eq: A_crystal} and the identity
\begin{equation}
\bm k\cdot (\bm q\times \hat{\bm z})\equiv \bm k* \bm q = k_x q_y-k_y q_x,
\end{equation}
one finds that, for any coupling with $\Delta\bm k_{ij}=\bm k_i-\bm k_j$ in the first reciprocal-lattice shell, the matrix elements of the paramagnetic Hamiltonian are
\begin{equation}
\langle i\vert \hat H_A^{(1)}\vert j\rangle
= -i\,\frac{e\hbar}{m^*c}\,\frac{\gamma}{2}\,\frac{(\bm k_j* \Delta\bm k_{ij})}{\vert \Delta\bm k_{ij}\vert ^2}\,e^{-i\Delta\bm k_{ij}\cdot \bm r_0}.
\end{equation}
An explicit evaluation performed with the above expressions for $\bm G_j$ and $\bm K$ gives
\begin{eqnarray}
\frac{\bm k_2 * \bm G_1}{G^2}&=& -\frac{1}{2\sqrt3},\\
\frac{\bm k_3 * \bm G_2}{G^2}&=& +\frac{1}{2\sqrt3},\\
\frac{\bm k_3 * (\bm G_3)}{G^2}&=& -\frac{1}{2\sqrt3}~.
\end{eqnarray}
Therefore, introducing the real amplitude
\begin{equation}
 t_A\equiv \frac{e\hbar}{m^*c}\,\frac{\gamma}{4\sqrt3}
=\frac{e\hbar}{m^*c}\,\frac{\beta\Phi_0}{8\pi\sqrt3},
\label{eq:tA}
\end{equation}
the $\hat H_A^{(1)}$ matrix in the basis \eqref{eq:three_states} at $\bm K$ is
\begin{equation}
H_{A,\rm para}(\bm r_0)=t_A\begin{pmatrix}
0 & i e^{-i\phi_1} & -i e^{-i\phi_2}\\
-i e^{i\phi_1} & 0 & i e^{i\phi_3}\\
 i e^{i\phi_2} & -i e^{-i\phi_3} & 0
\end{pmatrix}.
\label{eq:HA3K}
\end{equation}

The first-order in $\bm A$ degenerate perturbation theory on the doublet subspace is simply the projected operator
$H_{K}^{(1, \rm para)}\equiv \hat P H_{A,\rm para} \hat P$. We write $(1, \rm para)$ to indicate that the resulting matrix is the first order contribution in $\bm A$ coming from the paramagnetic term. 
Writing
\begin{equation}
H_{K}^{(1,\rm para)}=d_0^{(1)}\,\openone+d_x^{(1)}\sigma_x+d_y^{(1)}\sigma_y+m_{\rm p}^{(1)}\,\sigma_z,
\label{eq:proj_form}
\end{equation}
a long but straightforward calculation gives
\begin{align}
 d_0^{(1)}(\phi_1,\phi_2) &= \frac{t_A}{3}\Big[-\sin\phi_1+\sin\phi_2+\sin(\phi_1-\phi_2)\Big],\\
 d_x^{(1)}(\phi_1,\phi_2) &= \frac{t_A}{3}\Big[-\sin\phi_1+\sin\phi_2-2\sin(\phi_1-\phi_2)\Big],\\
 d_y^{(1)}(\phi_1,\phi_2) &= \frac{\sqrt3\,t_A}{3}\Big[\sin\phi_1+\sin\phi_2\Big],\\
 m_{\rm p}^{(1)}(\phi_1,\phi_2) &=-\frac{\sqrt3\,t_A}{3}\Big[\cos\phi_1+\cos\phi_2+\cos(\phi_1-\phi_2)\Big].
\label{eq:meven}
\end{align}
The resulting gap $\Delta_K$ in the Dirac spectrum is controlled by the $\sigma_z$ term: $\Delta_K=2\vert m_{\rm p}^{(1)}\vert $ at this order.
The terms $d_{x}$ and $d_{y}$ shift the Dirac point in momentum, while $d_0$ shifts the Dirac-point energy. 

\subsection{Magnetic perturbation at $K$: second order in $A$}
\subsubsection{Paramagnetic contribution}
We now calculate the second-order contribution of the paramagnetic term. The second-order paramagnetic contribution to the doublet Hamiltonian is given by
\begin{equation}
\begin{split}
H_{K}^{(2,\rm para)}&= \hat P H_{A,\rm para}(\bm r_0) \hat Q\,\frac{1}{E_d-QH_VQ}\,\hat Q H_{A,\rm para}(\bm r_0) \hat P\\
&=  - \frac{1}{3W}\,\hat P H_{A,\rm para}(\bm r_0)\hat Q H_{A,\rm para}(\bm r_0)\hat P,
\label{eq:SW}
\end{split}
\end{equation}
where $H_{A,\rm para}(\bm r_0)$ is inside the three-state manifold, $\hat P$ and $\hat Q$ are the matrix projecting respectively onto the doublet and the singlet, and we used the unperturbed energies $E_d=-W$ for the doublet and $E_s=+2W$ for the singlet, measured with respect to $E_0$.
By defining the doublet-singlet coupling amplitudes as
\begin{equation}
 v_\pm \equiv \langle u_\pm\vert H_{A,\rm para}(\bm r_0)\vert s\rangle,
\end{equation}
Eq.~\eqref{eq:SW} implies that the second-order correction is
\begin{equation}
\delta H_{K}^{(2)} = -\frac{1}{3W}
\begin{pmatrix}
\vert v_+\vert ^2 & v_+ v_-^*\\
 v_- v_+^* & \vert v_-\vert ^2
\end{pmatrix}.
\end{equation}
In particular, the correction to the Dirac mass induced by the second-order perturbation stemming from the paramagnetic Hamiltonian matrix $H_{A,\rm para}(\bm r_0)$ is
\begin{equation}
 m_{\rm p}^{(2)}(K)\equiv \frac{1}{2}\big[\delta H^{(2)}_{K}\big]_{11}-\frac{1}{2}\big[\delta H^{(2)}_{K}\big]_{22}= -\frac{1}{6W}\Big(\vert v_+\vert ^2-\vert v_-\vert ^2\Big).
\label{eq:modd_def}
\end{equation}
A direct evaluation using \eqref{eq:HA3K} and the explicit vectors $\vert s\rangle,\vert u_\pm\rangle$ gives the compact closed form
\begin{equation}
 m_{\rm p}^{(2)}(\phi_1,\phi_2) = -\frac{\sqrt3\,t_A^2}{27\,W}
\Big[4\sin\phi_1\sin\phi_2\sin(\phi_1-\phi_2)-\sin\phi_1+\sin\phi_2+\sin(\phi_1-\phi_2)\Big].
\label{eq:modd}
\end{equation}
Because the linear combination $-\sin\phi_1+\sin\phi_2+\sin(\phi_1-\phi_2)$ vanishes at first order in $\phi$, we have $m_{\rm p}^{(2)}=\mathcal O(\vert \bm r_0\vert ^3)$ for $\vert \bm r_0\vert \ll a_0$. 

\subsubsection{Diamagnetic contribution}
We now include the diamagnetic term, which is by itself second order in $\bm A$:
\begin{equation}
\hat H_A^{(2)}=\frac{e^2}{2m^*c^2}\,\bm A^2(\hat{\bm r}-\bm r_0).
\end{equation}
Within the first-shell approximation (i.e. retaining only Fourier components at $\pm \bm G_\ell$ and the uniform piece),
$A^2$ produces (i) a uniform energy shift and (ii) an additional cosine potential ``locked'' to the magnetic texture:
\begin{equation}
\hat H_A^{(2)}\;\approx\; \Delta E_{\rm dia}\,\mathbb{I} \;+\; 2t_{A^2}\sum_{\ell=1,3,5}\cos\!\big(\bm G_\ell\cdot(\hat{\bm r}-\bm r_0)\big),
\label{eq:HA2_firststar_final}
\end{equation}
where
\begin{equation}
t_{A^2}=\frac{e^2}{8m^*c^2G^2}\left(\frac{\beta\Phi_0}{2\pi}\right)^2,\qquad
\Delta E_{\rm dia}=6t_{A^2}.
\label{eq:WA2_Dedia}
\end{equation}
Higher-shell harmonics (e.g.\ $2\bm G_\ell$ and $\bm G_\ell-\bm G_{\ell'}$) are also generated by $\bm A^2$ but are omitted here
because they require enlarging the plane-wave truncation beyond the $K$-point three-state manifold. We obtain the correction to the doublet Hamiltonian introduced by the diamagnetic term as
\begin{equation}
H_{K}^{(2, \rm dia)} =d_0^{(2)}\,\mathbb{I}+d_x^{(2)}\,\sigma_x+d_y^{(2)}\,\sigma_y+m_{\rm d}^{(2)}\,\sigma_z\ ,
\end{equation}
with the explicit coefficients

\begin{gather}
d_0^{(2)}(\phi_1,\phi_2)
=-\frac{t_{A^2}}{3}\Big[\cos\phi_1+\cos\phi_2+\cos(\phi_1-\phi_2)\Big],
\label{eq:d0_A2}
\\
d_x^{(2)}(\phi_1,\phi_2)
=\frac{t_{A^2}}{3}\Big[-\cos\phi_1-\cos\phi_2+2\cos(\phi_1-\phi_2)\Big],\\ 
d_y^{(2)}(\phi_1,\phi_2)
=\frac{\sqrt3\,t_{A^2}}{3}\Big[\cos\phi_1-\cos\phi_2\Big],
\label{eq:dxdy_A2}
\\
m_{\rm d}^{(2)}(\phi_1,\phi_2)
=\frac{\sqrt3\,t_{A^2}}{3}\Big[\sin\phi_1-\sin\phi_2-\sin(\phi_1-\phi_2)\Big]~,
\label{eq:mA2_K}
\end{gather}
where the phases $\phi_j$ are defined in Eqs.~\eqref{eq:phis}. We stress that these contributions are second order in the external field, as shown in Eq.~\eqref{eq:WA2_Dedia}

\subsection{Final effective Hamiltonian}
We now provide the final expression of the Hamiltonian up to $\mathcal O(A^2)$. Since the derivation given above was done in valley $K$, we now comment on how to transition to valley $K'$. Repeating the same steps at $\bm K'=-\bm K$ (using the analogous three-plane-wave manifold at $\bm K'$) gives the following transformations on the Dirac velocity $\hbar v$ and shift phases $\phi_i$:
\begin{equation}
    \hbar v\big\vert _{K^\prime} = -\hbar v\big\vert _{K}~,\qquad \phi_i\big\vert _{K^\prime} = -\phi_i\big\vert _{K}~.
\end{equation}

Introduce the valley index $\tau=+1$ at $K$ and $\tau=-1$ at $K'$. The low-energy Hamiltonian near each valley, to linear order in
$\bm q$ and including (i) the $\bm k\cdot \bm p$ Dirac kinetic term, (ii) the direct $\mathcal O(A)$ projection of $\hat H_A^{(1)}$,
(iii) the $\mathcal O(A^2)$ SW correction from $\hat H_A^{(1)}$, and
(iv) the $\mathcal O(A^2)$ diamagnetic contribution \eqref{eq:HA2_firststar_final}, is
\begin{equation}
\begin{aligned}
H_{\tau}^{\rm eff}(\bm q)
&=\Big[(E_0-W)+\Delta E_{\rm dia}\Big]\mathbb{I}
+\tau\,\hbar v\,(q_y\sigma_x+q_x\sigma_y)
\\
&\quad
+\tau\Big[d_0^{(1)}\mathbb{I}+d_x^{(1)}\sigma_x+d_y^{(1)}\sigma_y\Big]
+m_{\rm p}^{(1)}\,\sigma_z
\\
&\quad
+\Big[d_0^{(2)}\mathbb{I}+d_x^{(2)}\sigma_x+d_y^{(2)}\sigma_y\Big]
+\tau\,m_{\rm d}^{(2)}\,\sigma_z
+\tau\,m_{\rm p}^{(2)}\,\sigma_z
\;+\;\mathcal O(q^2,A^3)~.
\label{eq:Heff_final_with_A2_SM}
\end{aligned}
\end{equation}
Here $d_{0,x,y}^{(1)}$ and $m_{\rm p}^{(1)}$ are the first-order [$\mathcal O(A)$] coefficients from the projection of $\hat H_A^{(1)}$, $m_{\rm p}^{(2)}$ is the valley-odd $\mathcal O(A^2)$ mass generated by the SW correction, and
$d_{0,x,y}^{(2)},m_{\rm d}^{(2)}$ are given in \eqref{eq:d0_A2}--\eqref{eq:mA2_K}.
The Dirac masses at the two valleys are, therefore,
\begin{equation}
m_{\tau}=m_{\rm p}^{(1)}+\tau\Big(m_{\rm d}^{(2)}+m_{\rm p}^{(2)}\Big)~.
\label{eq:final_masses}
\end{equation}

\subsection{Passing from $K$ to $K^\prime$}\label{appC}
Notice that the Hamiltonian obtained in Eq.~\eqref{eq:Heff_final_with_A2_SM} is written such that, in each valley, the basis of the Hilbert space is identified by the doublet obtained from the diagonalization of the crystal Hamiltonian. While the mathematical structure of the doublet is the same in valley $K$ and $K'$, its meaning is different since the plane-wave basis used to expand the starting Hamiltonian has a different physical meaning. In order to obtain our final continuum Hamiltonian, we must establish a common basis for the two valleys.

We now express the valley-$K'$ Hamiltonian in the same two-component (doublet) basis used at valley $K$.
This requires identifying the unitary transformation relating the doublet at $K'$ to the doublet at $K$.
At $K$, we use the three degenerate plane waves
\begin{equation}
\vert 1\rangle=\vert \bm K\rangle,\qquad \vert 2\rangle=\vert \bm K-\bm b_1\rangle,\qquad \vert 3\rangle=\vert \bm K-\bm b_2\rangle,
\end{equation}
and at $K'=-K$ we use
\begin{equation}
\vert 1'\rangle=\vert \bm K'\rangle,\qquad \vert 2'\rangle=\vert \bm K'+\bm b_1\rangle,\qquad \vert 3'\rangle=\vert \bm K'+\bm b_2\rangle.
\end{equation}
For spinless electrons, time reversal $\mathcal T$ acts as complex conjugation in the position representation and maps
plane waves according to $\mathcal T\vert \bm k\rangle=\vert -\bm k\rangle$. Therefore,
\begin{equation}
\mathcal T\vert 1\rangle=\vert 1'\rangle,\qquad \mathcal T\vert 2\rangle=\vert 2'\rangle,\qquad \mathcal T\vert 3\rangle=\vert 3'\rangle,
\label{eq:T_maps_triplet}
\end{equation}
i.e.\ $\mathcal T$ identifies the three-dimensional $K$-manifold with the three-dimensional $K'$-manifold.

In the $K$-triplet basis, the scalar potential $H_V^{(3)}$ is real and symmetric. Hence, its doublet eigenvectors can be chosen as
\begin{gather}
\vert u_+^{K}\rangle=\frac{1}{\sqrt3}\Big(\vert 1\rangle+\omega \vert 2\rangle+\omega^2\vert 3\rangle\Big),\\
\vert u_-^{K}\rangle=\frac{1}{\sqrt3}\Big(\vert 1\rangle+\omega^2 \vert 2\rangle+\omega\vert 3\rangle\Big).
\label{eq:uK_def}
\end{gather}
Analogously, in the $K'$-triplet basis, we define
\begin{gather}
\vert u_+^{K'}\rangle=\frac{1}{\sqrt3}\Big(\vert 1'\rangle+\omega \vert 2'\rangle+\omega^2\vert 3'\rangle\Big),\\
\vert u_-^{K'}\rangle=\frac{1}{\sqrt3}\Big(\vert 1'\rangle+\omega^2 \vert 2'\rangle+\omega\vert 3'\rangle\Big).
\label{eq:uKp_def}
\end{gather}
Using \eqref{eq:T_maps_triplet} and the antiunitarity of $\mathcal T$ (which complex-conjugates coefficients), we obtain
\begin{align}
\mathcal T\vert u_+^{K}\rangle
&=\frac{1}{\sqrt3}\Big(\mathcal T\vert 1\rangle+\omega^* \mathcal T\vert 2\rangle+(\omega^2)^*\mathcal T\vert 3\rangle\Big)\\
&=\frac{1}{\sqrt3}\Big(\vert 1'\rangle+\omega^2\vert 2'\rangle+\omega\vert 3'\rangle\Big)
=\vert u_-^{K'}\rangle,\\
\mathcal T\vert u_-^{K}\rangle
&=\frac{1}{\sqrt3}\Big(\vert 1'\rangle+\omega\vert 2'\rangle+\omega^2\vert 3'\rangle\Big)
=\vert u_+^{K'}\rangle.
\end{align}
Thus, under time reversal, the two doublet components are exchanged between valleys:
\begin{equation}
\mathcal T\vert u_+^{K}\rangle = \vert u_-^{K'}\rangle,\qquad
\mathcal T\vert u_-^{K}\rangle = \vert u_+^{K'}\rangle.
\label{eq:T_swaps_doublet}
\end{equation}

Let the ordered doublet bases be
\begin{equation}
\mathcal B_K=\big(\vert u_+^{K}\rangle,\ \vert u_-^{K}\rangle\big),\qquad
\mathcal B_{K'}=\big(\vert u_+^{K'}\rangle,\ \vert u_-^{K'}\rangle\big).
\end{equation}
Eq.~\eqref{eq:T_swaps_doublet} implies that the identification of the $K'$ doublet with the $K$ doublet is given,
up to overall phase conventions, by exchanging the two components. At the level of two-component pseudospinors this is implemented by
\begin{equation}
U=\sigma_x=
\begin{pmatrix}
0 & 1\\
1 & 0
\end{pmatrix},
\qquad
\begin{pmatrix}
c_+^{(K)}\\ c_-^{(K)}
\end{pmatrix}
=
U
\begin{pmatrix}
c_+^{(K')}\\ c_-^{(K')}
\end{pmatrix}.
\label{eq:U_sigma_x}
\end{equation}
Let $H_{K'}$ be the effective $2\times 2$ Hamiltonian written in the $K'$ doublet basis $\mathcal B_{K'}$.
That Hamiltonian, expressed in the $K$ doublet basis $\mathcal B_K$, is
\begin{equation}
H_{K'}^{(K\text{-basis})}=U\,H_{K'}^{(K'\text{-basis})}\,U^\dagger,\qquad U=\sigma_x.
\label{eq:HKp_transform}
\end{equation}
Finally, the full Hamiltonian obtained in Eq.~\eqref{eq:Heff_final_with_A2_SM}, written in the $K$-basis, and apart from irrelevant constant energy shifts, takes the following transparent form:
\begin{equation}
\begin{aligned}
H_{\tau}^{\rm eff}(\bm q)\Big\vert _{K\text{-basis}}
&=\,\hbar v\,(\tau q_y\sigma_x+q_x\sigma_y)+(\tau m_{\rm p}^{(1)}+m_{\rm d}^{(2)} + m_{\rm p}^{(2)})\,\sigma_z
\\
&\quad
+\Big[\tau d_0^{(1)}\mathbb{I}+ (\tau d_x^{(1)}+ d_x^{(2)})\sigma_x+(d_y^{(1)}+\tau d_y^{(2)})\sigma_y\Big]
\\
&\quad+\;\mathcal O(q^2,A^3).
\end{aligned}
\end{equation}
This Hamiltonian corresponds to the one given in the main text in Eq.~\eqref{eq:Heff_final_with_A2} with $m_{\rm H} = m_{\rm p}^{(1)}$, $m_{\rm S} = m_{\rm d}^{(2)} + m_{\rm p}^{(2)}$, $q_{0,x} = -(\tau d_x^{(1)}+ d_x^{(2)})$, and $q_{0,y} = -(d_y^{(1)}+ \tau d_y^{(2)})$.

\section{Different magnetic field pattern}
In this Section, we provide the topological phase diagram for a different choice of PMF with respect to that specified in the main text in Eqs.~(\ref{V_crystal}) and (\ref{B_crystal}). Even in this case we find extended regions of the phase diagram with non-zero Chern number, proving that the particular choice of PMF used in the main text is not essential for obtaining non-trivial band topology.

\begin{figure}[h]
    \centering
    \begin{overpic}[width=0.333\linewidth]{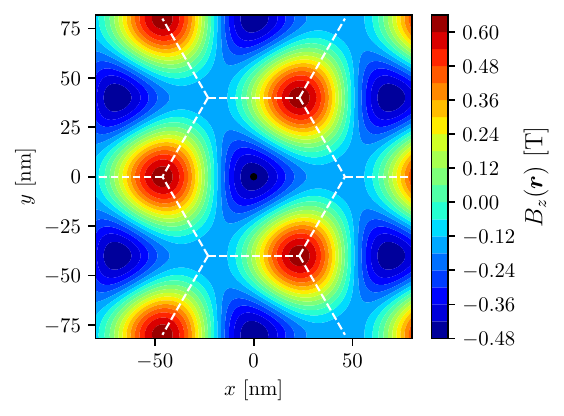}
        \put(0,72){(a)}
    \end{overpic}%
    \begin{overpic}[width=0.333\linewidth]{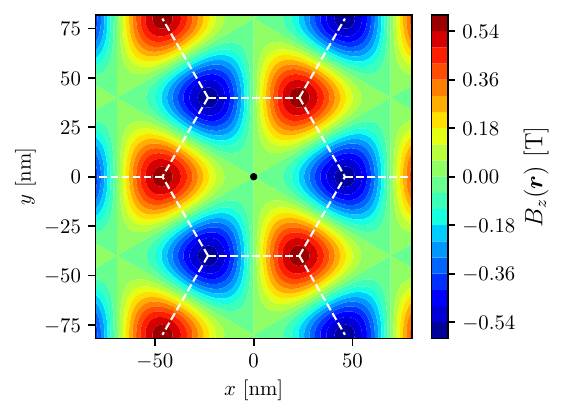}
        \put(0,72){(b)}
    \end{overpic}%
    \begin{overpic}[width=0.333\linewidth]{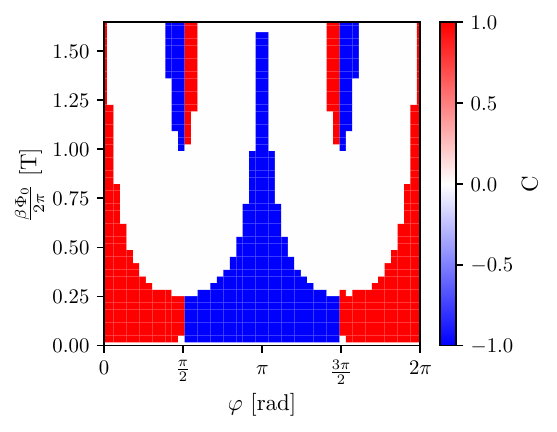}
        \put(0,72){(c)}
        \put(50,45){\color{white} $\bm r_0$}
    \end{overpic}
    \caption{PMF $B_{z}(\bm r)$ as obtained from Eq.~\eqref{B_crystal_supp} with $\varphi = \frac{\pi}{4}$, in panel (a), and $\varphi = \frac{\pi}{2}$, in panel (b). The topological phase diagram is shown in panel (c).  ${\cal C}$ is the Chern number of the lowest band $\epsilon_{\bm k,\lambda=1}$ plotted as a function of the magnetic field intensity $\beta \Phi_0 /2\pi$ and the phase $\varphi$. The other parameters are fixed as in the main text: $a_0 = 80\ {\rm nm}$ and $W = 1.57\ {\rm meV}$.}
    \label{fig:SM2}
\end{figure}

Using the same crystal potential $V(\bm r)$ as in the main text [Eq.~\eqref{V_crystal}], we now consider a PMF of the form
\begin{equation}\label{B_crystal_supp}
    B_z(\bm r)= \frac{\beta}{2\pi}\Phi_0 \sum_{\ell=1,3,5}
    \cos (\bm G_\ell \cdot \bm r \ + \varphi)~,
\end{equation}

which is periodic with the lattice periodicity
and differs from Eq.~(\ref{B_crystal}) by the fact that all reciprocal-lattice harmonic, identified by $\bm G_\ell$, are now shifted by the same phase $\varphi$. We emphasize that the effect of this phase is not equivalent to the global spatial shift $\bm r_0$ considered in the main text. Rather, $\varphi$ can be interpreted as shifting each harmonic by a distinct vector $\bm r_{0,\ell}$, defined through $\bm G_\ell\cdot\bm r_{0,\ell}\equiv \varphi$. The resulting PMF patterns obtained from Eq.~\eqref{B_crystal_supp} are shown in Fig.~\ref{fig:SM2}(a) and (b)
for two different values of $\varphi$.

They preserve the threefold rotational symmetry, while differing substantially from those analyzed in the main text, see Fig.~\ref{fig0}(b) and (c).
In Fig.~\ref{fig:SM2}(c), we show the Chern number of the lowest band as a function of the magnetic-field amplitude $\beta \Phi_0 /2\pi$ and of the phase $\varphi$,
demonstrating the presence of regions with nonzero Chern number.

\end{document}